\documentclass[preprint2,epsfig]{aastex}

\usepackage{float}
\usepackage{fixltx2e}
\usepackage{natbib}
\title{
Grouping Normal Type Ia Supernovae by UV to \\
Optical Color Differences}

\author{Peter~A.~Milne\altaffilmark{1},
Peter~J.~Brown\altaffilmark{2},
Peter~W. A.~Roming\altaffilmark{3}, \\
Filomena Bufano\altaffilmark{4}, 
Neil Gehrels\altaffilmark{5}}
\altaffiltext{1}{University of Arizona, Steward Observatory,
                    933 N. Cherry Ave., Tucson, AZ 85719, USA.}
\altaffiltext{2}{George P. and Cynthia Woods Mitchell Institute for Fundamental Physics \& Astronomy,
Texas A. \& M. University, Department of Physics and Astronomy,
4242 TAMU, College Station, TX 77843, USA;
pbrown@physics.tamu.edu}
\altaffiltext{3}{Southwest Research Corporation}
\altaffiltext{4}{Universidad Andres Bello, Departmento de Cincias Fisicas, 
Avda. Republica 220, Santiago, Chile}
\altaffiltext{5}{NASA-Goddard Space Flight Center, Astrophysics Science
Division, Codes 660.1 \& 662, Greenbelt, MD 20771, USA.}

\newcommand{\totNormSNe}{23 }
\newcommand{\newNormSNe}{11 }

\begin{document}
\begin{abstract}

Observations of many type Ia supernovae for multiple epochs per object
with the {\it Swift} UVOT instrument has revealed that there
exists order to differences in the UV-optical colors of optically normal
supernovae. We examine UV-optical color curves for \totNormSNe SNe Ia, dividing
the SNe into four groups, finding that roughly one-third of ``NUV-blue" SNe~Ia
have bluer UV-optical colors than the larger ``NUV-red" group.
Two minor groups are recognized, ``MUV-blue" and ``irregular" SNe~Ia. While the 
latter group is concluded to be a subset of the NUV-red group, containing the 
SNe with the broadest optical peaks, the ``MUV-blue" group is concluded to be a 
distinct group. Separating into the groups and accounting for the
time evolution of the UV-optical colors, lowers the scatter in
two NUV-optical colors (e.g. $u-v$ and $uvw1-v$) to the level of the scatter
in $b-v$. This finding is promising for extending the cosmological utilization of
SNe~Ia into the NUV. We generate spectrophotometry of 33 SNe~Ia and
determine the correct grouping for each. We argue that there is a fundamental 
spectral difference in the 2900--3500$\AA$ wavelength range, a region suggested to be dominated by
absorption from iron-peak elements. The NUV-blue SNe~Ia feature less absorption 
than the NUV-red SNe~Ia. We show that all NUV-blue SNe~Ia in this
sample also show evidence of unburned carbon in optical spectra, whereas only
one NUV-red SN~Ia features that absorption line. Every NUV-blue event also exhibits
a low gradient of the SiII $\lambda$6355\AA\ absorption feature.
Many NUV-red events also exhibit a low gradient, perhaps suggestive that NUV-blue events
are a subset of the larger LVG group.

\end{abstract}

\section{INTRODUCTION}

Type Ia supernovae (SNe Ia) are luminous events that synthesize an 
appreciable fraction of the iron-peak elements in the universe (Iwamoto et al. 1999), 
and their high and relatively homogeneous peak luminosities have allowed them to 
be utilized to measure large distances. SN Ia distance estimates 
have revealed that the expansion of the universe is accelerating 
(Riess et al. 1998; Perlmutter et al. 1999). 
There is considerable interest in extending the current utilization of 
SNe Ia, both by better understanding the SN Ia event and by widening the 
rest-frame wavelength range which can be utilized.  
The ultraviolet wavelength (UV) range, including the range covered 
by the $U$-band, bears tremendous potential towards the better understanding 
of SNe Ia, but it also poses significant challenges.\footnote{In this work, 
we treat the UV wavelength range to be emission shortward of 4000$\AA$, 
NUV to be emission between the 2500 -- 4000$\AA$ ($u$ and $uvw1$ filters) 
and MUV to be emission between 1500 -- 2500$\AA$ ($uvm2$ and $uvw2$ filters).} 

Theoretically, the UV wavelength range is useful to 
better understand the explosive nucleosynthesis of the SN event. 
Iron-peak elements strongly affect this wavelength range through 
the related effects of line-blanketing and line-blocking (Jeffery 
et al. 1992. Sauer et al. 2008, Hachinger et al. 2013). 
In particular, observations made at early epochs are a sensitive 
probe of the composition of the outer layers of the SN ejecta. 
This is important for the determination of the progression of the 
burning front through the outer layers of the ejecta during a 
suggested delayed detonation phase. 
The flux in the UV part of the spectrum of a SN Ia is shaped mostly
by metal lines, in particular FeII and III, TiII, CrII, CoII and NiII, 
with contributions from SiII and MgII. 
These ions have a large number of strong as well as weak lines in the UV.
UV photons are actively absorbed in these lines, and because of the high
velocity of the SN ejecta the forest of lines becomes a ``blanket":
different lines in different parts of the ejecta overlap because
of velocity shear, leading to the effective blocking of the flux.
Once photons are absorbed, they are most likely to escape only if
they are re-emitted in red transitions, as at red (optical) wavelengths
the total line opacity is much less.
Since metal line opacity is quite high,
the UV spectra are therefore good tracers of metal abundance in the
outer layers of the ejecta, and different temperatures or abundance ratios
can lead to different UV colours (e.g. Walker et al 2012).
On the other hand, in the outermost layers, UV opacity may not be so high because
the density is low.
In this case the strongest metal lines are the only active ones. These
are typically in the optical. It is therefore possible for optical photons
to be absorbed and re-emitted in the UV in a process called reverse fluorescence.
Because of the low UV opacity in these outermost layers, reverse fluoresced
photons are essential in determining the emerging UV spectrum (e.g. Mazzali 2000).
Hence, observations made in the UV combined with knowledge of the velocity of the 
expansion of the ejecta from spectra, permit detailed probing of the radiation 
transport on the SN ejecta. 

Cosmologically, the 
wavelength range covered by the $U$-band is potentially useful 
for distance studies. Extending a ground-based study of high redshift 
SNe Ia beyond z$\sim$0.5 requires either observations made in 
observer-frame NIR (rest-frame optical), or in the observer-frame 
optical (rest-frame UV). To 
explore the potential of the $U$-band, ground-based SN surveys have 
included $U$-band photometry in their recent campaigns of low-redshift 
SNe. These 
campaigns have reported that there is considerable scatter 
in the photometry, at a level that exceeds what is seen in the optical 
filters. This scatter persists after all efforts to account for 
observational issues. Jha, Riess \& Kirshner (2007) presented 
$UBVRI$ light curves 
of 44 SNe Ia, but found that their multi-light curve fitting algorithm 
had lower overall scatter without including the $U$-band, than it did 
with the $U$-band. 
 
One critical element for SN Ia cosmology is to determine whether SN Ia 
emission evolves with redshift. Thought to be the thermonuclear explosion of a 
carbon-oxygen white dwarf (hereafter CO WD) that accretes material from a companion, 
evolution of emission might be caused by metallicity, 
differences between low and high redshift progenitor CO WDs or by 
different companions to the CO WD between low and high redshift 
binary systems. Since modeling suggests that the UV wavelength range is 
a sensitive probe of the nucleosynthetic products of the SN explosion 
(i.e. the composition of the ejecta, Sauer et al. 2008; Lentz et al. 2000), 
a number of recent studies have compared rest-frame UV emission from 
multiple SNe Ia, trying to quantify the variations between events. 
The rest-frame UV emission for low-$z$ samples have been obtained from 
SN Ia observations made with IUE (Capellaro et al. 1995, Panagia et al. 2003), 
from observations made with HST (Jeffery et al. 1992, 
see Foley, Filippenko \& Jha 2008 for a listing of all IUE/HST UV spectra through 2007), 
from high-$z$ samples with Keck (Ellis et al. 2008, Foley et al. 2012c), 
or from HST observations of recent SNe Ia detected at very early epochs by the 
Palomar Transient Factory (Cooke et al. 2011, Maguire et al. 2012, Hachinger et al. 2013) 
or other recent SNe~Ia (2009ig: Foley et al. 2012a, 2011iv: Foley et al. 2012b,  
2011by: Foley \& Kirshner 2013). 
These studies have reported some differences between the UV emission 
from high-$z$ versus low-$z$ samples, and that the variations seen in the UV wavelength 
range within a sample 
exceeds the variations seen in the optical wavelength range. 
Combined with the difficulties experienced with the ground-based $U$-band 
SN~Ia light curve surveys, the UV wavelength range has been 
considered a problematic wavelength range for cosmological studies. 

\begin{figure*}[t]
\epsscale{1.6} \plotone{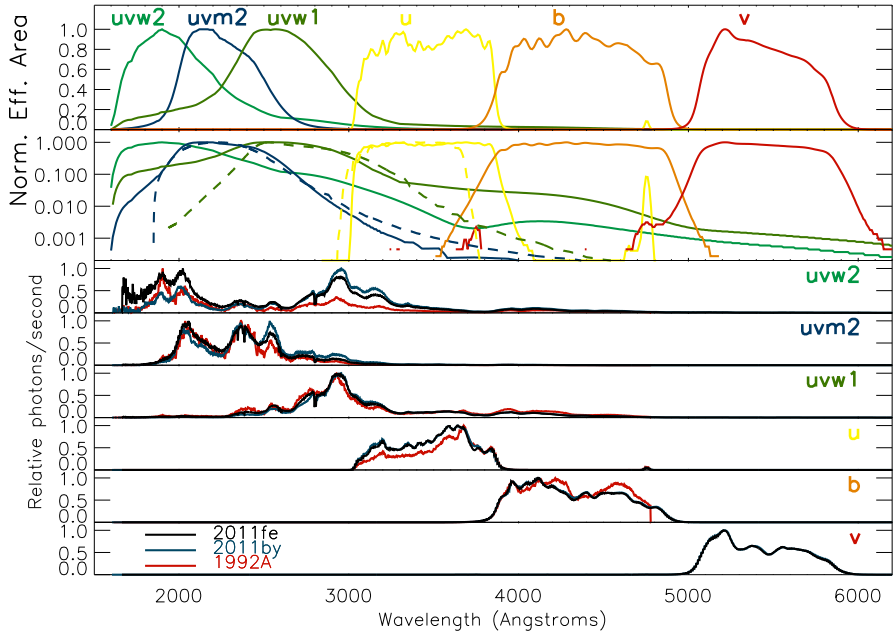}
\caption{The six UVOT filter transmission curves folded through the
{\it HST}/CTIO spectrum of SN 1992A (Kirshner et al. 1993) in red, the 
UVOT spectrum of SN~2009ig (Foley et el. 2012a), and the HST spectrum of 
SN~2011iv (Foley et al. 2012b). The transmission
curves for the six UVOT filters are shown in the top two panels on linear and 
logarithmic scales. HST F220W, F250W and F330W are plotted with dashed lines on 
the logarithmic panel for comparison. The lower 
panels show the three spectra folded through the 
UVOT transmission curves (solid lines).}
\label{transcurves}
\end{figure*}

The UVOT instrument (Roming et al. 2000, 2005) on the {\it Swift} mission 
(Gehrels et al. 2004) possesses 
the ability to quickly schedule observations of a new supernova, 
increasing the chances of observing the important early epochs, 
prior to peak light. 
The 30cm UVOT instrument employs 3 UV filters ($uvw1, uvm2, uvw2$) and 
three optical filters ($u,b,v$), and SNe are typically observed with 
all six filters at all epochs. The transmission curves for each filter are shown in 
Figure 1, folded through three UV/optical spectra of SNe~Ia. Brown et al. (2009) presented 
photometry from the first  2.5 years of UVOT observations, 2133 total observations 
of 25 SNe, of which 17 are SNe Ia. Whereas HST cameras have the advantage of superior 
S/N photometry and spectroscopy for the SNe Ia that are 
observed, the advantage for UVOT observations are superior 
sampling of SNe Ia, both in number of events and in the number of 
observations for a single event. The photometry presented in 
Brown et al. (2009) was studied for light curve properties in 
Milne et al. 2010 (hereafter M10) and for 
absolute magnitudes in Brown et al. 2010 (hereafter B10). 
M10 identified UV-optical color differences between the group
of normal SNe~Ia and the groups of 
subluminous and SN 2002cx-like SNe Ia. The definition of
normal used in M10, which is the same definition that will be used in this work, is 
a SN Ia with a B-band peak decline rate less than 1.6 (e.g. $\Delta m_{15}(B)$ $\leq$ 1.6), 
and not exhibiting optical spectral features similar to SN~2002cx. This group comprises the 
majority of SNe Ia, and is generally the group utilized for cosmological distance studies. 
Within this group of normal SNe Ia, the findings of M10 and B10 have not wholly matched the 
trends derived from ground-based optical and $U$-band surveys, nor the intepretations 
of the HST and IUE UV spectra, as the initial sample of UVOT UV light curves appeared 
more homogeneous than expected. If the UV light curves are intrinsically homogeneous, or,  
barring that, can be organized into 
distinct groups and exhibit low intrinsic scatter within those groups, 
the UV wavelength range would prove useful for cosmological applications. 

In this work, we add two more years of SN Ia UVOT observations, 
to the M10/B10 collection, showing how the larger sample permits a better 
understanding of the range of UV light curves possible within the group 
of normal SNe~Ia (Section 2). We present a more quantitative investigation into the 
color evolution of sub-classes we recognize within ``normal" SNe~Ia. 
We will not present color curves for subluminous SNe Ia, as those will be 
presented in a separate work (Milne et al., in preparation).
In Section 3, we compare UVOT photometry with spectrophotometry extracted from HST 
spectra of SNe~Ia. In Section 4, we search for a correlation between blue UV-optical colors 
and optical parameters, such as the spectral signature of unburned carbon. 
In Section 5, we treat extinction and how extinction estimation needs to adapt to new 
findings. New UVOT photometry is presented in the appendix. 

In the first collective studies of UVOT SN Ia photometry, M10 and B10  
found that the color evolution of the UVOT-$u$ and
UV filters relative to the optical ($v$) band was both dramatic and
relatively homogeneous within the sub-class of normal SNe Ia. The
colors become rapidly bluer until $\sim$6 days pre-peak in the
$B$ filter, then abruptly become redder until $\sim$20 days.\footnote{We use 
``BPEAK" to denote the date of the maximum brightness in the $B$ filter.}
There was a lone significant outlier to that trend, SN~2008Q, which followed the 
same color-curve shape but was bluer
at all early epochs than the other normal SNe Ia, seemingly offset from the 
other normal events. The absolute magnitudes of the light curve peaks
of the UVOT-$u$ light curves were able to be fit with a
linear luminosity-width relation (LWR). The scatter about that $u$ band
LWR was similar to what was seen in the $b$ and $v$ filters. The
LWR for the $uvw1_{rc}$ filter had somewhat larger scatter and the
$uvw2_{rc}$ and $uvm2$ filters had large scatter.\footnote{$uvw1_{rc}$  and
$uvw2_{rc}$ refer to ``red-tail corrected" $uvw1$ and $uvw2$ filters.
See B10 for details of the method used to try to account for
contamination of those filters from longer wavelength emission.}
However, the
peak pseudocolors were not correlated with peak-width for the NUV-optical
colors, a finding different than what has been seen in the $B-V$ colors
(the Phillips peak relation; Phillips et al. 1999), or in ground-based $u-V$ colors
(Folatelli et al. 2010).

Collectively, with the exception of SN~2008Q,
the UVOT data-set of normal SNe~Ia suggested that there
is a high level of homogeneity in the $U$-band wavelength range, but
significant variations at shorter wavelengths. These findings are
promising for the cosmological utilization of the NUV emission from SNe~Ia, 
as it suggests that the bright $U$ band can be successfully used to measure
distances with low intrinsic scatter. These findings are also interesting
for probing the explosion physics of SN Ia events, as the dramatic
color change with time and the intrinsic variations at the shorter wavelengths
provide benchmarks for SN Ia modeling.

Since the B10 and M10 presentations of UVOT SN Ia photometry, the final 
photometric reductions for \newNormSNe more normal SNe Ia have been completed. 
The photometry for SNe 2008hv, 2009ig, and SNF2008-0514 
were presented in Brown et al. (2012a), photometry for 2009dc was 
presented in Silverman et al. (2011), and photometry for SN~2011fe was 
presented in Brown et al. (2012b). The photometry for SN~2009an, SN~2009cz,
SN~2010cr(=PTF10fps), PTF09dnl, PTF10icb, SN~2010gn(=PTF10mwb), SN~2011by and SN~2011iv     
is presented in Appendix A.\footnote{SNe~2010cr(=PTF10fps) and 2011iv are narrow-peaked 
SNe~Ia included for comparisons of HST spectrophotometry with UVOT photometry.} All photometry 
follow the method outlined in Brown et al. (2009), which uses post-SN 
template images to remove non-SN counts from the data, and incorporate the 
updated zeropoints and time-dependent sensitivity corrections of Breeveld et al. (2011). 
Results of fitting functions to the data will be presented in a future work, 
this work will concentrate on color curves. 

UVOT photometry was supplemented with ground-based photometry for SNe~2005df and 
2011fe. Australia National University data for 2005df was published in M10, 
included because UVOT only observed that SN with the UV filters. Stritzinger et al. (2010) 
published UBV photometry as part of an optical/NIR study of SN~2006dd and three other 
SNe~Ia in the host galaxy, NGC~1316. That data was used, as the UVOT optical data at peak 
suffered severe coincidence loss. SN~2011fe also suffered severe coincidence loss 
at optical peak, so $B$ and $V$ band data was used from Richmond \& Smith (2012).

The reddening estimations are shown in Table \ref{extinct}, and are based upon 
the sum of the Milky Way Galaxy (MWG) reddening and the host reddening. 
The MWG reddening was obtained 
from Schlafly \& Finkbeiner (2011) via NED and the host reddening was obtained by the 
Phillips peak pseudocolor relation (Phillips 1999), the Lira tail color relation (Lira 1995) 
and/or from ground-based studies of these SNe Ia, 
the host reddening being the remainder after subtracting 
MWG reddening. For the majority of this work, the reddening estimates are used only for 
sample selection. In Section 5, the complications of treating extinction in light of 
newly recognized groupings is discussed.  

Red-leak corrections, as explained in B10, were not employed in this work. 
K-corrections are estimated using a collection of UV-optical spectra in Appendix B, but 
were not employed for the color curves. The lack of UV spectra to provide 
adequate time coverage of all the major and minor groups that we have identified in this paper is 
the reason that neither correction is implemented for the color curves.
The two filters that suffer the largest effect of red-leak, in terms of the fraction of 
photons estimated to originate from the wavelength range of the adjacent redder filter, 
are the $uvw1$ and $uvw2$ filters. As will be shown, the color curves that include these 
filters are not simply tracings of the color curves from the adjacent redder filter. This 
suggests that the most red-leak 
affected filters still provide unique information about photons in their central wavelength 
range. Nonetheless, readers are cautioned to avoid treating any of the filters as 
box functions about a central wavelength.  

\begin{table*}
\scriptsize
\vspace{-9mm}
\caption{Peak width and extinction estimates for the SN Ia Sample}
\begin{tabular}{l|c|ccccc|c}
\hline
\hline
SN & $\Delta m_{15}(B)$ & E(B-V)$_{MWG}^{a}$ & E(B-V)$_{host}(peak)$ & 
E(B-V)$_{host}$(tail) & E(B-V)$_{host}$(lit)$^{b}$ & 
Ref.$^{c}$ & E(B-V)$_{TOT}$(used) \\
   &  [mag] & [mag]  & [mag]  & [mag]  & [mag]  &  & [mag] \\
(1) & (2) & (3) & (4) & (5) & (6) & (7) \\
\hline
\multicolumn{6}{c}{Normal}  \\
\hline
2005cf & 1.07 & 0.087 &   -0.01 $\pm$ 0.05 &  0.16 $\pm$ 0.04 &
         0.11 $\pm$ 0.03 & 1,3 & 0.21 \\
2005df & 1.20 & 0.026 &    0.13 $\pm$ 0.15 &  0.02 $\pm$ 0.10 &
         ---  & 1 &  0.05 \\
2006dd & 1.34 & 0.019 &    0.05 $\pm$ 0.01 &  --- $\pm$ --- &
         0.04$\pm$ 0.01  & 2,9 &  0.06 \\
2006dm & 1.54 & 0.034 &    0.05 $\pm$ 0.08 &  0.05 $\pm$ 0.01 &
         0.04$\pm$ 0.09  & 1,3 & 0.09  \\
2006ej & 1.39 & 0.031 &    0.18 $\pm$ 0.12 &  0.05 $\pm$ 0.01 &
         0.03 $\pm$ 0.09 & 1,3 & 0.09 \\
2007af & 1.22 & 0.034 &    0.17 $\pm$ 0.08 &  0.05 $\pm$ 0.01 &
         0.15 $\pm$ 0.04 & 1,3 & 0.19 \\
2007cq & 1.04 & 0.095 &    0.10 $\pm$ 0.12 &  ---  &
         --- & 1  & 0.21 \\
2007cv & 1.31 & 0.062 &    0.14 $\pm$ 0.08 &  ---  &
         --- & 1  & 0.21 \\
2007gi & 1.37 & 0.021 &  --- &  --- &
         0.16 $\pm$ 0.05 & 2,3 & 0.21 \\
2007sr & 1.16 & 0.041 &    0.12 $\pm$ 0.05 &  --- &
         0.11 $\pm$ 0.08 & 2,3 & 0.16 \\
2008Q & 1.40 & 0.073 &     0.02 $\pm$ 0.08 &  --- &
         --- & 1,8  &  0.10 \\
snf08-0514 & 1.20 & 0.030 & -0.09 $\pm$ 0.07 & -0.07 $\pm$ 0.16 &
         --- & 8  &  0.03 \\
2008ec & 1.08 & 0.061 &  0.16 $\pm$ 0.08 &  0.20 $\pm$ 0.15 &
         --- & 1  &  0.25 \\
2008hv & 0.95 & 0.029 &  -0.12 $\pm$ 0.06 & -0.010 $\pm$ 0.13 &
         --- & 5 &  0.03 \\
2009an & 1.20 & 0.016 &   --- &  --- &
         --- & 5 &  --- \\
2009cz & 0.99 & 0.023 &   0.04 $\pm$ 0.01 &  --- &
         --- & ---$^{d}$  &  0.07 \\
2009dc & 0.72 & 0.062 &  -0.29 $\pm$ 0.10 &  -0.01 $\pm$ 0.08 &
         --- & 5 &  0.17 \\
2009ig & 0.70 & 0.028 &   0.22 $\pm$ 0.20 &  0.09 $\pm$ 0.06 &
         0.00 $\pm$ 0.01 & 5,6 &  0.12 \\
09dnl & 0.98 & --- & --- & --- & 
       --- & 10 & --- \\ 
2010gn$^{f}$ & 1.19 & 0.027 & --- & --- & 
           0.05$^{e}$ & 10 & 0.05 \\ 
10icb & 1.09 & 0.011 & 0.06 $\pm$ 0.04 & --- &  
          0.08$^{e}$ & 10 & 0.08 \\
2011by & 1.14 & 0.012 & --- & --- & 
            -0.04$^{g}$ & 12 & 0.015 \\
2011fe & 1.21 & 0.008 &    --- &  --- &
         --- & 7  &  0.025 \\
\hline
\multicolumn{6}{c}{Narrow-peaked}  \\
\hline
2010cr$^{f}$ & 1.79 & 0.031 & --- & --- &
              --- & 10 & --- \\
2011iv & 1.69 & 0.010 & --- &  --- &  0.01 & 11  &  0.01 \\
\hline
\end{tabular}
\begin{tabular}{l}
$^{a}$ MWG extinction estimates were obtained from Schlafly \& Finkbeiner 2011 
              via NED. \\
$^{b}$ E(B-V)$_{host}$(lit) refers to host galaxy estimates from sources other \\
than Brown et al. (2010) [B10] and Milne et al. (2010) [M10] \\
$^{c}$ References: (1) B10 and references therein, (2) M10 and references therein, \\
(3) Wang et al. (2009b), (4) Folatelli et al. (2012), (5) Hicken et al. (2009) \\
(6) Foley et al. (2011), (7) Richmond \& Smith (2012), (8) Ganeshalingam et al. \\
(2010), (9) Stritzinger et al. (2010), (10) Maguire et al. (2012), \\
(11) Foley et al. 2012, (12) Foley \& Kirshner 2013 \\
$^{d}$ E(B-V) value for 2009cz derived from current UVOT data. \\
$^{e}$ E(B-V) values for SN 10icb and SN~10mwb are total values from Maguire et al. 2012,
using the Folatelli et al. (2012) \\
peak color relation. $\Delta m_{15}(B)$ = 1.96(1/s -1) +1.07 stretch to  $\Delta m_{15}(B)$
conversion used (Perlmutter et al. 1997). \\
$^{f}$ PTF10fps=SN~2010cr, PTF10mwb=SN~2010gn. \\
$^{g}$ Derived from Silverman, Ganeshalingam \& Filippenko (2011) B-V using 
Folatelli et al. (2012) peak color relation. \\
\end{tabular}
\label{extinct}
\end{table*}

\begin{figure*}[t]
\epsscale{1.75} \plotone{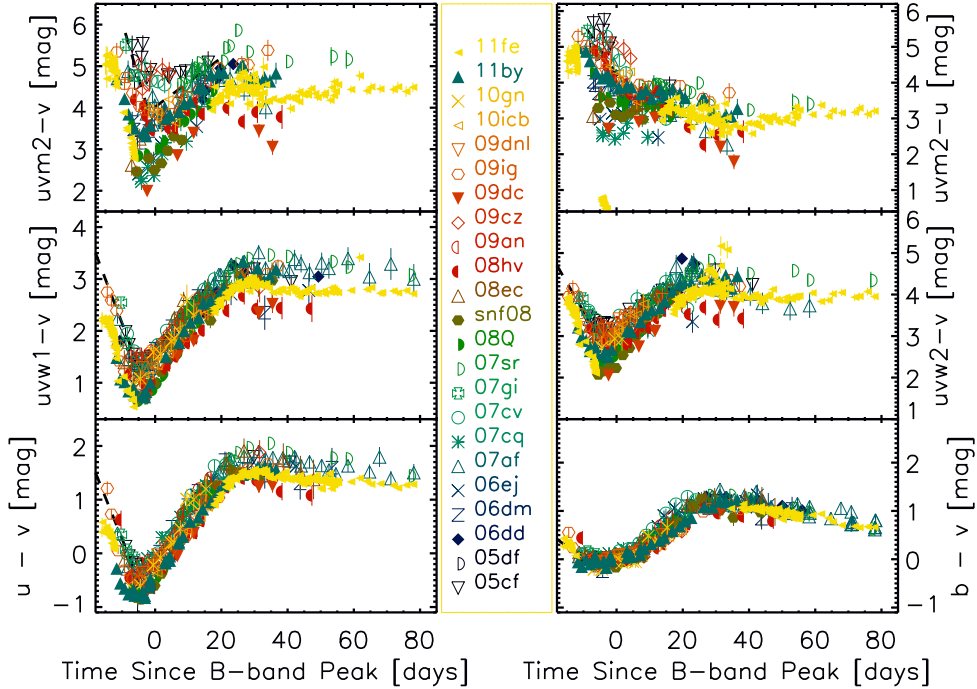}
\caption{Colors of 23 normal SNe~Ia relative to the $v$ band.
The slope of the NUV-$v$ color changes are steeper, cover a larger 
total change and switch to a reddening trend more abruptly than the 
$b-v$ colors.}
\label{col_N}
\end{figure*}

\begin{figure*}[t]
\epsscale{1.8} \plotone{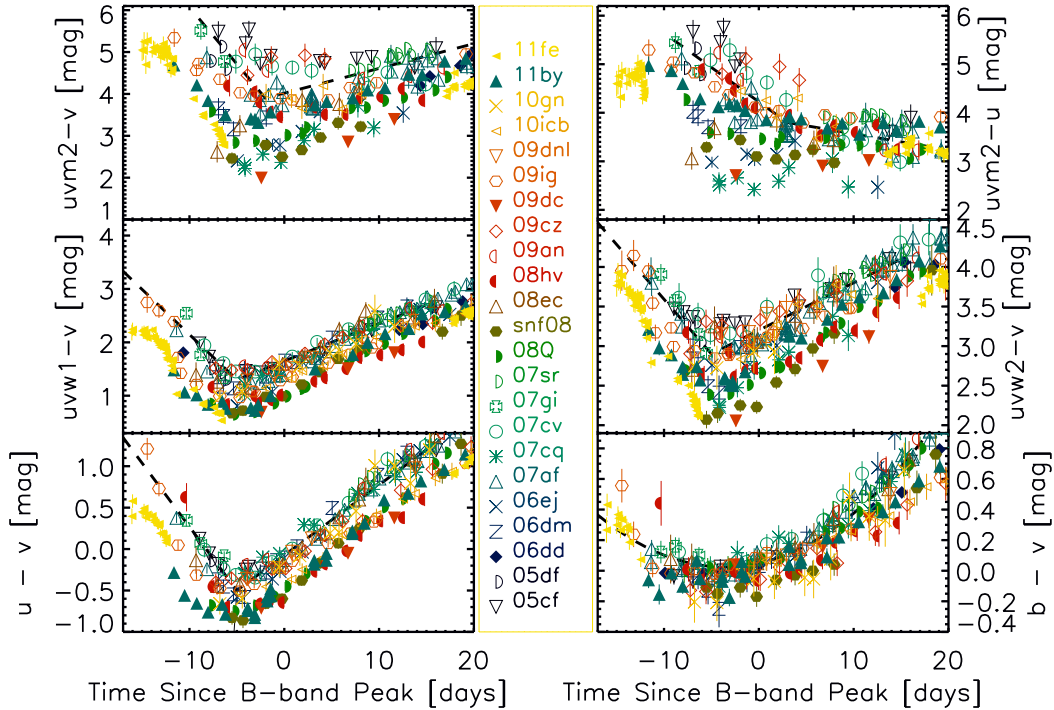}
\caption{Near-peak colors of 23 normal SNe~Ia compared to the 
$v$ band. The ``NUV-blue" group is shown 
with filled symbols, the ``NUV-red" group with open symbols, the 
MUV-blue group is shown with plusses and crosses. The $uvm2-u$ colors 
are shown in the upper right panel to permit comparison of MUV versus 
NUV colors.}
\label{col_N_zoom}
\end{figure*}

\begin{figure*}[t]
\epsscale{1.80} \plotone{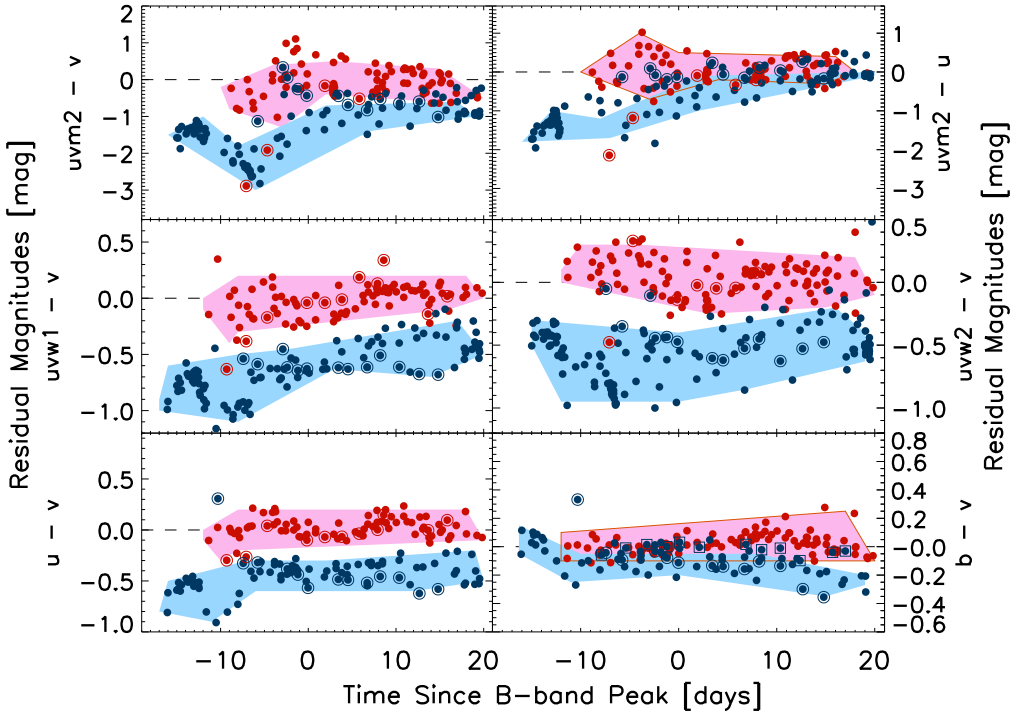}
\caption{Residuals of near-peak NUV-optical colors to ``U" and ``V" shaped 
curves shown in Figure \ref{col_N_zoom} for a subset of the SN~Ia sample. 
SNe~Ia are grouped as NUV-blue (blue) or NUV-red (red). Due to 
tendencies that somewhat defy the overall grouping, the data for the 
NUV-blue SN~2008hv is circled as is the data for the NUV-red SN~2008ec, 
while the NUV-blue SN~2008Q is boxed. See text for discussion.}
\label{color_residsA}
\end{figure*}   

\begin{figure*}[t]
\epsscale{1.80} \plotone{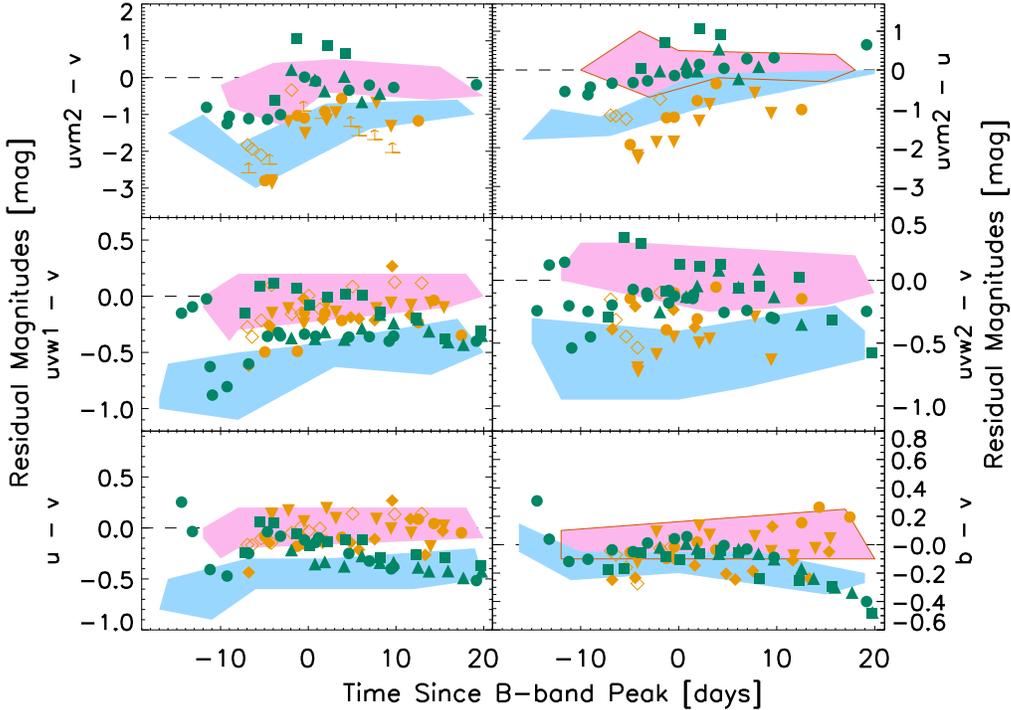}
\caption{Same as Figure \ref{color_residsA}, but showing the data for the 
minor MUV-blue (orange, 2006ej: filled circles, 2007cq: filled inverted triangles, 
2010gn: filled diamonds, 2006dm: open diamonds) and 
irregular (green, 2009ig: filled circles, 2009cz: filled squares, 10icb: filled triangles) 
groups. Observations of SN~2010gn yielded only 
upper limits for $uvm2$, shown in the $uvm2-v$ upper left panel.
Presented as 
residuals, the characteristic features of the two minor groups are evident. 
MUV-blue: similar to NUV-red in redder filters, but similar to NUV-blue in
the bluer filters. Irregular: following the blue edge of NUV-red at peak, 
but evolving to as blue as NUV-blue by +10 days.}
\label{color_residsB}
\end{figure*}

\begin{figure*}[t]
\epsscale{1.8} \plotone{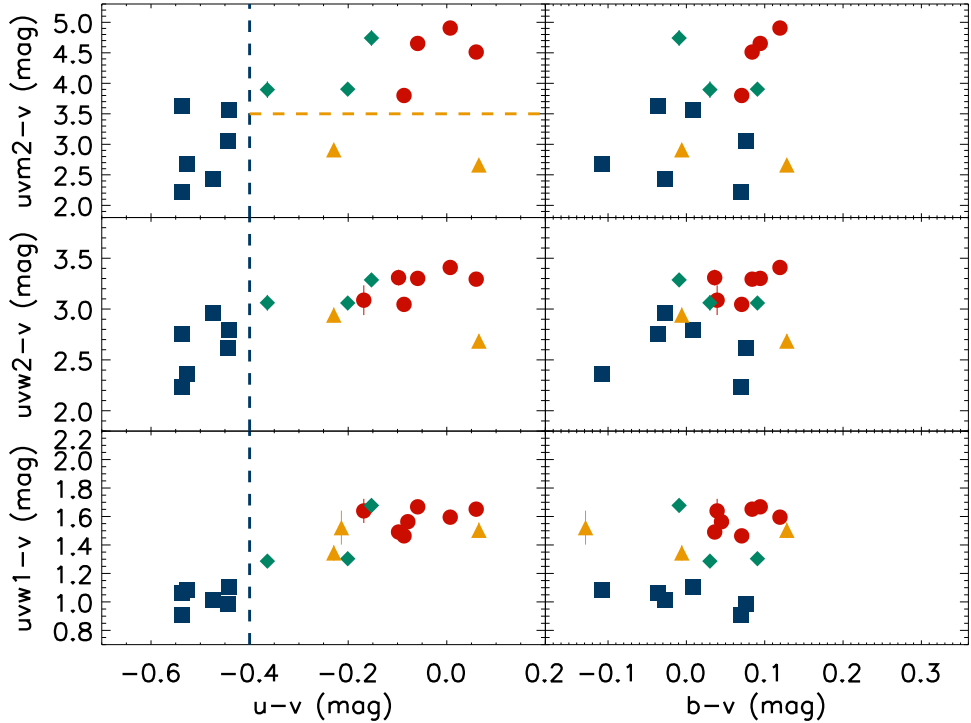}
\caption{Color-color plots of 19 SNe~Ia determined at BPEAK. The 
color at BPEAK was determined by a linear fit to the data with 
-4 $\leq$ $t$ $\leq$ +15 days relative to BPEAK. 
SNe~Ia are grouped as NUV-blue (blue-square),
NUV-red (red-circle), MUV-blue (orange-triangle) and irregular (green-diamond). 
The blue, verticle dashed line shows a provisional cut line in $u-v$ to 
separate NUV-blue from NUV-red/irregular, and the orange, horizontal dashed 
line shows a provisional cut line in $uvm2-v$ to separate MUV-blue from 
NUV-red/irregular. 
}
\label{color_color_offset}
\end{figure*}

\begin{figure*}[t]
\epsscale{1.8} \plotone{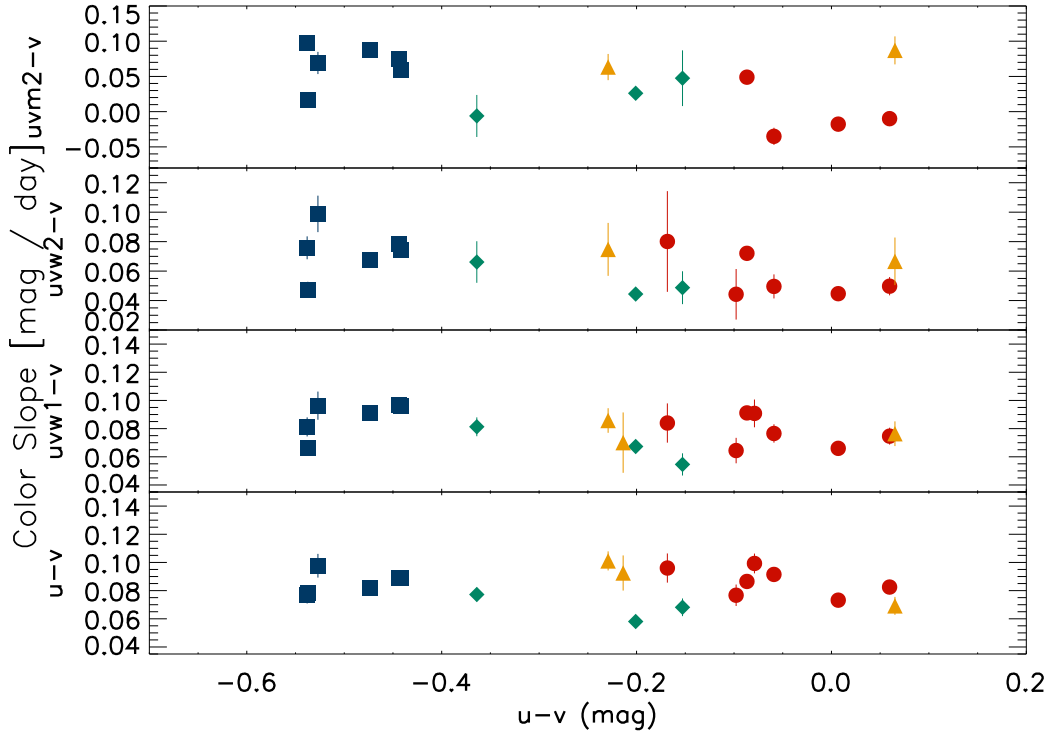}
\caption{Slope of color evolution of 19 SNe~Ia during near-peak 
epoch. The linear fits are the same as used in Figure \ref{color_color_offset}. 
SNe~Ia are grouped as NUV-blue (blue-square),
NUV-red (red-circle), MUV-blue (orange-triangle) and 
irregular (green-diamond).}
\label{color_color_slope}
\end{figure*}

\begin{figure*}[t]
\epsscale{1.8} \plotone{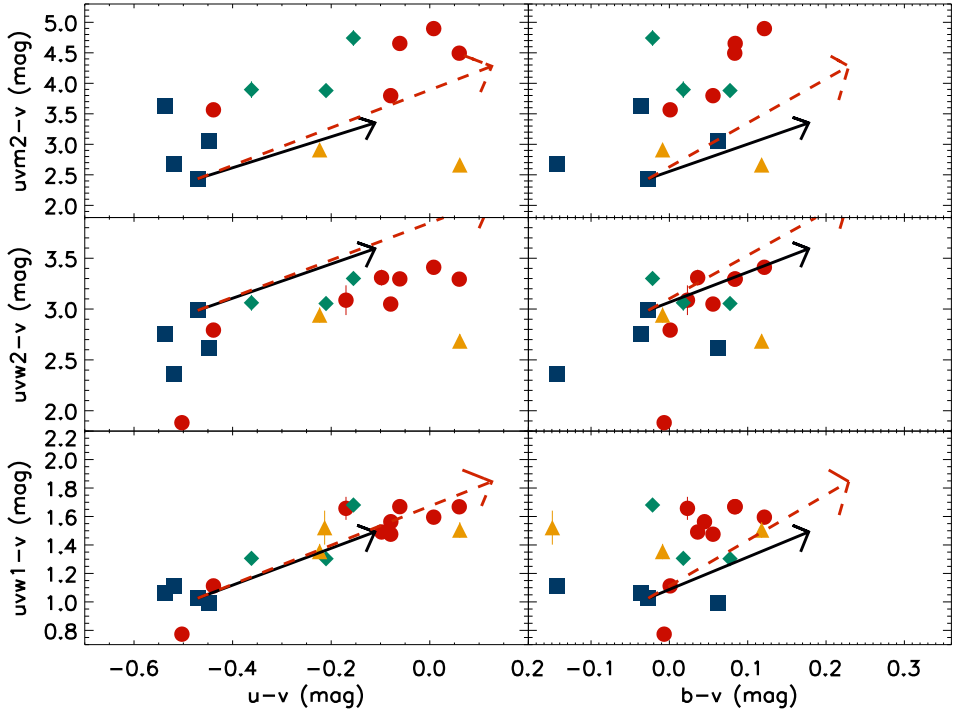}
\caption{Color-color plots of UVOT SN Ia sample with reddening vectors. MWG dust is 
shown with black solid arrows and CSLMC dust with red dashed arrows. SN symbols are 
the same as for Figure \ref{color_color_offset}.}
\label{redv}
\end{figure*}

\begin{figure*}[t]
\epsscale{1.8} \plotone{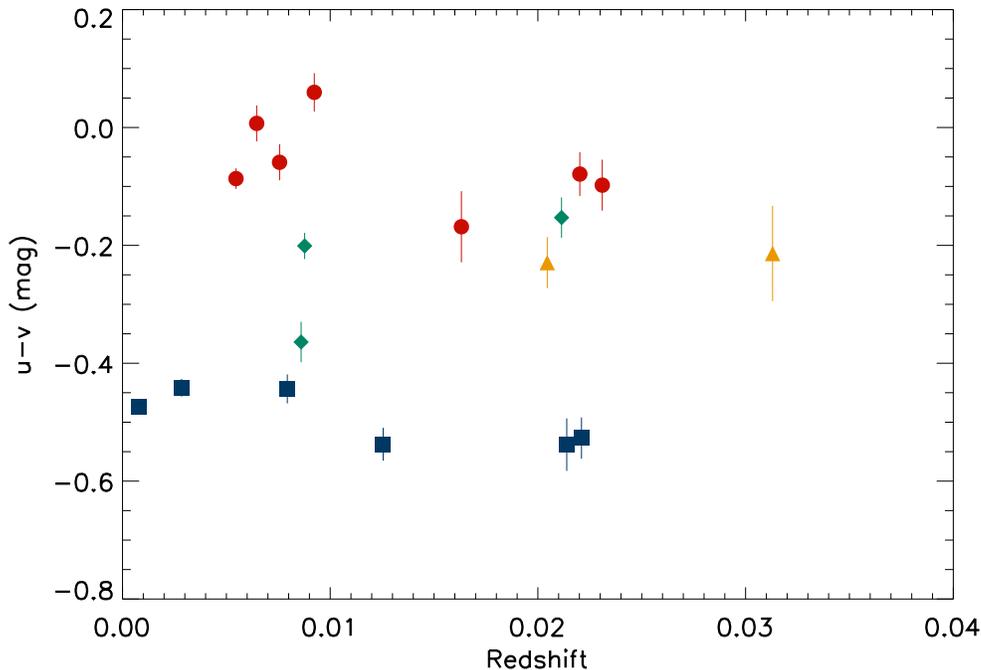}
\caption{$u$-$v$ colors at BPEAK plotted versus redshift. 
SNe~Ia are grouped as NUV-blue (blue-square),
NUV-red (red-circle), MUV-blue (orange-triangle) and irregular (green-diamond).
There is no apparent correlation of colors with redshift between the groups.}
\label{color_vs_redshift}
\end{figure*}

\begin{figure*}[t]
\epsscale{1.8} \plotone{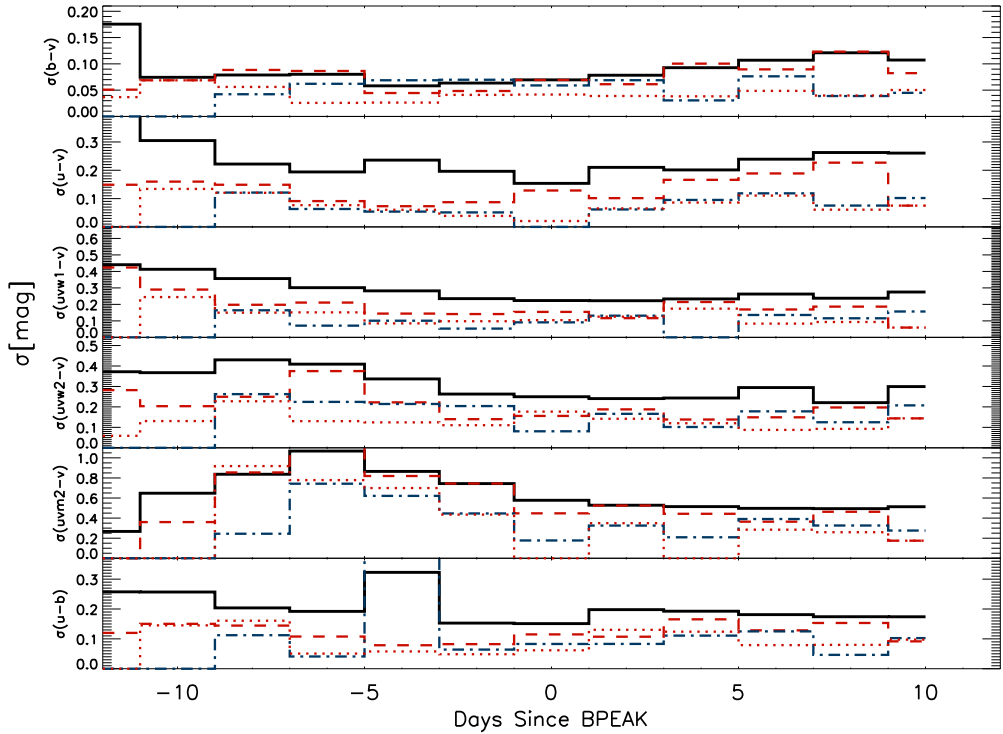}
\caption{Scatter of SN~Ia color curves from mean color curves as a 
function of epoch. 
The black solid curve shows the scatter for all SNe~Ia, while 
blue dot-dashed shows NUV-blue events. The red dashed groups 
NUV-red, MUV-blue and irregular groups, while the red dotted shows just the 
NUV-red group. The MUV-blue events are combined with the NUV-blue events 
for the $uvw2-v$ and $uvm2-v$ filters. Bins with fewer than 4 data points 
have been set to zero.}
\label{color_scatter}
\end{figure*}  

\begin{figure*}[t]
\epsscale{1.8} \plotone{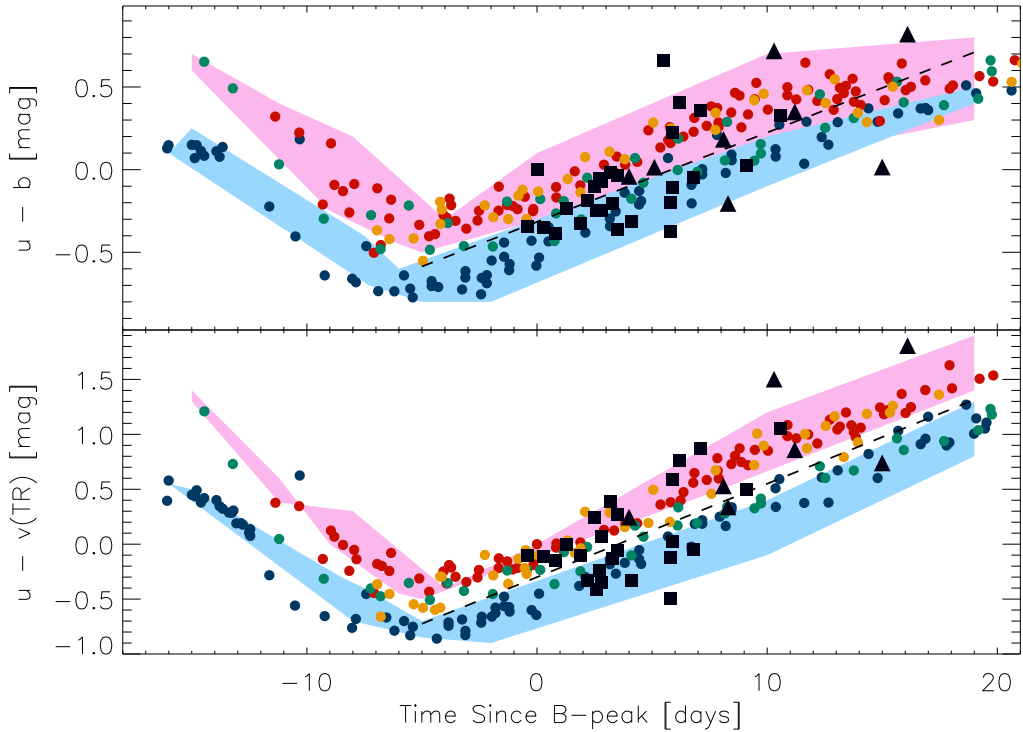}
\caption{$u-b$ and $u-v(TR)$ spectrophotometry derived from HST observations of 
SNe Ia compared with UVOT photometry (small filled circles). 
The HST spectra are from Maguire et al. (2012). The u-v(TR5500) 
colors approximate a $v$ band but with truncation at 5500\AA\.  
The u-v(TR) colors have been offset by 0.7 mag as an $S$-correction to match the 
UVOT $u-v$ photometry. Filled squares are SNe in the Maguire et al. 2012 study, 
filled triangles are SNe in the Foley et al. 2010 study. The dashed lines are linear 
fits to the UVOT NUV-red SN Ia photometry, offset by 0.1 mag and 0.2 mag, respectively 
to quantify separations between NUV-red and -blue. 
The range of UVOT photometry of NUV-red and -blue SNe~Ia is shown as red and blue 
shaded regions.}
\label{u_bv_cooke_foley}
\end{figure*}

\begin{figure*}[t]
\epsscale{1.8} \plotone{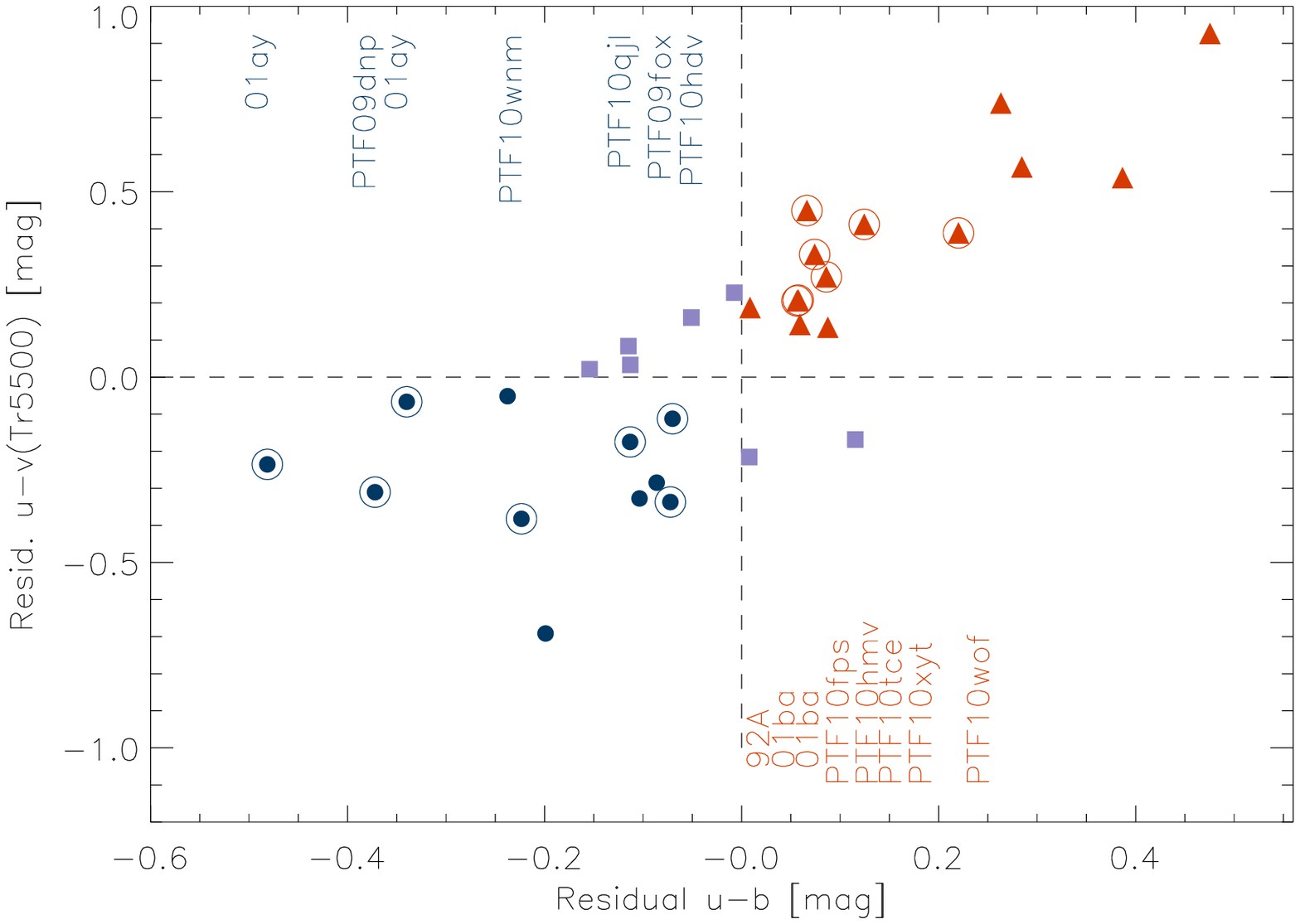}
\caption{Residuals of the HST spectrophotometry to the linear fits shown in Figure 
\ref{u_bv_cooke_foley}, with $u-b$ versus $u-v(TR5500)$. 
NUV-blue events (filled blue circles) 
are defined to have negative residuals in both colors, while NUV-red events 
(filled red triangles) are defined to have positive residuals in both colors. 
Events with mixed residuals (filled purple squares) are considered undefined. 
The NUV-blue and -red events used in the Figure \ref{hst_compspec} spectral 
comparisons are circled and labeled.}
\label{hst_color_color}
\end{figure*}

\section{UVOT PHOTOMETRY}

\subsection{Normal SN Ia UVOT Colors} 

The color curves of the normal SNe Ia are shown in Figure \ref{col_N}. 
We include only SNe Ia for which the total MWG and host galaxy reddening, E(B-V), is estimated to 
be less than 0.25 mag, but we apply no extinction correction. 
The basic features of the color curves 
are the same as shown in M10.\footnote{SN~2005am was considered a 
normal SN~Ia in M10, but has since been re-categorized as a narrow-peaked SN~Ia. It will 
be included in a study of narrow-peaked SNe~Ia (Milne et al., in preparation.} 
The magnitude 
of the color changes are larger for the NUV-$v$ colors than for the 
$b-v$, due to a steeper slope both during the initial blueward 
evolution and the subsequent redward evolution. The transition from 
blueward to redward is much more abrupt in the NUV-$v$ colors than 
for the $b-v$ color curves. We characterize the abrupt transitions of the 
NUV-$v$ color curves as ``V" shaped, and the more gradual transitions of the 
$b-v$ color curves as ``U" shaped. 
The slopes of the color evolution of the $u-v$ and $uvw1-v$ color 
curves are 0.08 mag d$^{-1}$ and 0.09 mag d$^{-1}$, respectively.
This means that when spectra are 
normalized by the V band wavelength range, the NUV emission can 
drop by 20\% in 3 days. This potential complication must be recognized 
when combining spectra into a mean spectrum or when comparing spectra 
between low-$z$ and high-$z$ samples.

The near-peak epochs are highlighted in Figure \ref{col_N_zoom}. The important 
addition from M10 is that there are now 7 SNe Ia with a high NUV/optical 
ratio (SNe 2006dd, 2008Q, SNF20080514-002, SNe 2008hv, 2009dc, 2011by and 2011fe),  
where in M10, only SN 2008Q exhibited that tendency. Throughout 
this paper, we will refer to this group as the ``NUV-blue" SNe Ia, and 
the larger group as ``NUV-red". 
The separation between the two groups far exceeds the scatter of 
each group relative to a mean color evolution. This can be 
quantified by fitting a straight line to each group during the 
-5 to +20 days epoch and determining the scatter about the mean curve. 
The scatter is 0.15 mag in the $u-v$ color 
for the NUV-red group and 0.10 mag for the NUV-blue group. By contrast, the 
average color curves of the 
two groups are separated by 0.44 magnitudes, far larger than the scatter 
about either line fit. This color difference suggests that it is critical 
to determine the NUV-blue or NUV-red grouping when studying $U/u$ band 
emission from SNe Ia, especially for standard candle applications. 
The same SNe Ia that exhibit blue $u-v$ colors also exhibit blue 
$uvw1-v$ colors, suggesting that this dichotomy is not related 
to $K$-corrections as would be created by a single abrupt drop in the 
NUV spectrum.

``V" and "U" shaped lines are also shown in Figure \ref{col_N_zoom}, 
representing the mean color evolution of the NUV-red group. 
In Figure \ref{color_residsA} we display the residuals of the colors 
curves relative to those lines for a subset of the overall sample 
(eliminating MUV-blue and irregular events until Figure 5). 
The NUV-blue and NUV-red events are 
plotted in blue and red, respectively, where the determination is 
based upon the $u-v$ and $uvw1-v$ color curves. The shaded regions represent a 
crude outer-boundary for each group. 
The NUV-blue nature is exhibited in all panels, but with much larger 
separation in the $u-v$, $uvw1-v$ and $uvw2-v$ colors, 
with the NUV-blue group separate from the NUV-red group at all epochs.
Although many of the NUV-blue
events are along the blue edge of the $b-v$ distribution, that trend is contrasted
by SN 2008Q, which is quite average relative to the NUV-red events.
There are two notable anomalies in the early-epoch data, where the NUV-blue SN~2008hv is 
initially red, and the NUV-red SN~2008ec that is initially blue.

In Figure \ref{color_residsB}, we show the remaining SNe~Ia and 
introduce two additional minor groups; ``irregular" events and ``MUV-blue" events. 
For clarity, we plot only the shaded regions from Figure \ref{color_residsA}, rather than 
the NUV-blue and -red data. The irregular events are  where at least one of 
the color curves evolves such a way that there is a deviation from 
the NUV-red group for some portion of the 
near-peak epoch. The MUV-blue events appear as NUV-red events 
for the $u-v$ and $uvw1-v$ color curves, but are blue in the $uvm2-v$ 
and $uvm2-u$ color evolution. 
The MUV-blue events, SN~2006ej, 2006dm, 2007cq and 2010gn(=PTF10mwb) are indistinguishable 
from the NUV-red events for the $u-v$ and $uvw1-v$ color curves, but 
trace the NUV-blue events for the $uvm2-v$ and $uvm2-u$ color curves. 
The irregular events appear along the blue edge of the NUV-red group for the 
week of $B$-band peak in the $u-v$ to $uvm2-v$ color curves, but the dominant 
characteristic is a different shape of the color evolution. By +15 days the irregulars are 
as blue as the NUV-blue SNe~Ia in all filters. 
SN~2009cz appears redder in the $uvw1-v$, $uvw2-v$ and $uvm2-v$ 
color curves than the other irregulars, but the slopes of 
the color evolution is a solid match to the irregular group. We note that 
SNe~2009ig and 2009cz dramatically redden a few days pre-peak, appearing to end a period of 
bluer NUV-optical emission during the -15 to -3 day epoch.

The trends seen in the color curves can be better visualized by deriving the 
colors of each SN at the time of $B$-band maximum. The t(BPEAK) 
color is determined by fitting a straight line to the -4 day to +10 day 
color evolution. Resulting color-color plots are shown in Figure 
\ref{color_color_offset}. The right panels show a high level of overlap between
the four groups, suggesting that the mean ($b-v$)$_ {BPEAK}$ color of NUV-blue events 
would be bluer than the mean color of any of the other three groups, but 
the ($b-v$)$_{BPEAK}$ color would be a poor determinant for group membership. By 
contrast, the ($u-v$)$_{BPEAK}$ color shows a clear separation between 
NUV-blue and NUV-red events in all three color-color plots. A provisional definition 
of NUV-blue could be ($u-v$)$_{BPEAK}$ $\leq$ -0.4 mag, as shown in the vertical dashed line 
of Figure \ref{color_color_offset}. 
The MUV-blue events are mixed with the NUV-red and irregular events in 
the ($uvw1-v$)$_{BPEAK}$ color, but are bluer in the $uvm2-v$ BPEAK color, as shown by a 
horizontal dashed line at ($uvm2-v$)$_{BPEAK}$ $\leq$ 3.5 mag. The irregular group tends 
toward the blue-edge of the NUV-red distribution, appearing as an extension of the NUV-red group. 
Figure \ref{color_color_slope} shows the slopes of the same fits, plotted versus 
the ($u-v$)$_{BPEAK}$. The slopes of the NUV-blue group are within the range of the 
NUV-red group, confirming that the primary difference 
between the two major groups is color offset, rather than slopes. The 
irregular group tends towards shallow slopes and blue $b-v$ peak colors, 
while the $uvm2-v$ color evolves 
rapidly for the MUV-blue group, perhaps another distinguishing feature of that 
group.

One potential explanation for the differences of NUV-optical colors could be 
that NUV-red events are instrinsically NUV-blue events, but suffer additional extinction. 
This possibility will be discussed more in Section 5, but in Figure \ref{redv}, 
we show reddening vectors that give the direction of changes on a color-color plot that 
would be caused by either Milky Way Galaxy (MWG) dust or LMC dust 
including circumstellar scattering (see Goobar 2005 and B10).  While the general 
color distribution is similar to that of a dust-reddened population,  the colors of 
individual objects in multiple filters do not appear consistent with reddening being the 
primary difference.  In other words, if one finds the reddening law direction and 
magnitude separating two SNe for a given pair of colors, that reddening correction 
would usually fail to match up the two SNe in a different color combination.  
Another 
potential explanation for the differences of NUV-optical colors could be that it is 
an observational effect caused by different redshifts in the sample. As no 
$K$-corrections have been applied in this study, the colors shown retain any effects 
of redshift. However, Figure \ref{color_vs_redshift} shows the $u$-$v$ colors at 
BPEAK plotted versus redshift. There are members 
of all four groups on both the left and right sides of the figure, suggesting that 
these groupings are not an artifact of no $K$-corrections.

\subsection{Scatter about Mean Color Curves}

Determination of whether categorization of the UV colors lowers 
the scatter within the collection of color curves to $\sim$0.1 mag is important for the 
cosmological utilization of the UV wavelength range. 
To probe the remaining scatter in colors after organizing the SNe~Ia into 
groups, we bin the data into two-day bins and determine mean colors for each 
bin. We then calculate the standard deviation of the data relative to the 
mean color. Figure \ref{color_scatter} shows the standard devation as a function 
of epoch for six colors, where bins with fewer than 4 data points have been set to zero. 
The scatter of the $b-v$ color (upper panel) is 
considered the comparison, as the rest-frame $B$ and $V$ filters are the basis of 
SN~Ia cosmology. The scatter is $\sim$0.08 mag near-peak, increasing at early 
and late epochs, in part due to having applied no stretch correction. 
Phillips et al. (1999) and Folatelli et al. (2010) have developed a peak-color correction, 
which we have not applied. In principle, 
either correction would lower the scatter to lower than 0.08 mag. 

The $u-v$ scatter is near 0.2 mag when all SNe~Ia are included and groupings ignored, 
becoming larger at early and late epochs. This is consistent with studies that have 
treated all normal SNe~Ia as one group. 
Separating the SNe~Ia into two groups, NUV-blue (blue dot-dashed) 
and all others (red dashed), leads to a much lower scatter (on the order of 0.13 mag). 
Excluding irregular and MUV-blue events further lowers the scatter, albeit by a 
small amount. Similar improvement occurs in the $uvw1-v$ and $u-b$ color evolution, when the 
NUV-blue events are separated from the NUV-red events. The scatter is considerable for the 
$uvw2-v$ and $uvm2-v$ color curves of the NUV-red SNe, remaining high even after separating 
out other groups. By contrast, for the NUV-blue SNe all color curves exhibit scatter on the 
order of 0.1 mag during the near-peak epoch, excepting the $uvm2-v$ color curves. 
Collectively, this suggests that after sorting into groups, 
the $u$ and $uvw1$ are promising wavelength 
ranges to contribute light curves to cosmological distance studies using SNe~Ia.  
This contribution could be the direct use of rest-frame $U/u$ band light curves of high-$z$ 
SNe Ia, or through minimizing scatter in the optical colors by distinguishing NUV-red from 
NUV-blue events. By contrast, the current sorting is unable to lower the scatter to useful 
levels of the shorter wavelength-optical color curves. 

\begin{figure*}[t]
\epsscale{1.8} \plotone{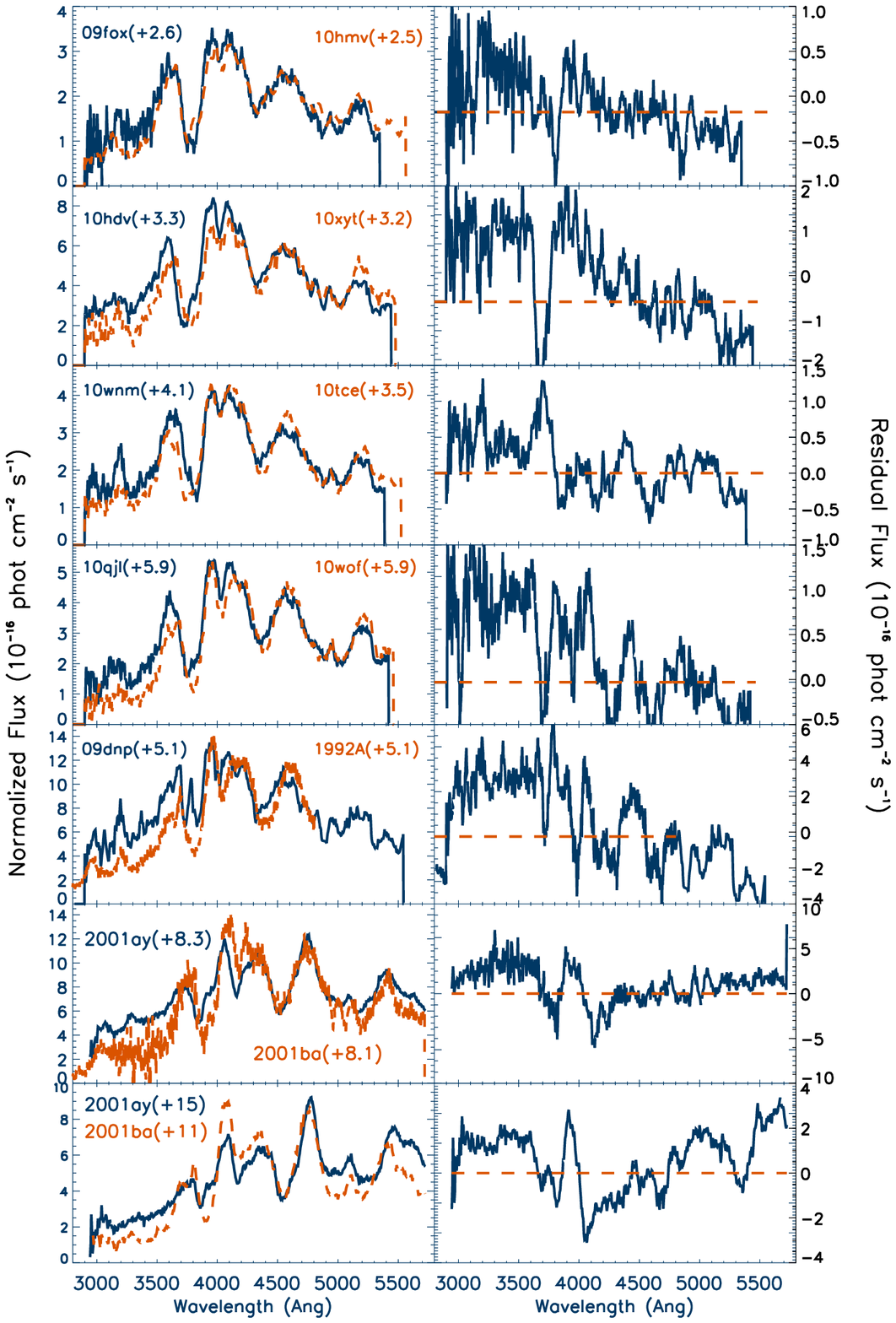}
\vspace{-2cm}
\caption{
Comparison HST UV-optical spectra of five pairs of NUV-blue and NUV-red 
SNe Ia at early epochs (upper 5 panels), and SN~2001ay (NUV-blue or irregular) 
versus SN~2001ba (NUV-red) at 2 later epochs(lower 2 panels). 
Solid blue lines are NUV-blue events, red-dashed lines 
are NUV-red events. The spectra are normalized to the overlap in the 
4000 -- 5000$\AA$ wavelength range, or to the red edge of the spectrum. 
The right panels show the residual of the 
NUV-blue minus NUV-red, in all cases showing a block excess shortward of 3600$\AA$.}
\label{hst_compspec}
\end{figure*}

\section{HST SPECTROPHOTOMETRY COMPARISONS}

The UVOT color curve relations presented and explored in the previous section, 
lead to interest in understanding the spectral nature of those differences, 
through comparisons of UV spectra of normal SNe~Ia. 
Ideally, UV-optical spectra of UVOT SNe~Ia could be matched by epoch and 
compared. However, there are too few such spectra for a complete comparison. 
An alternative is to utilize the existing collection of UV spectra to 
produce $u-v$ spectrophotometry, folding spectra through the UVOT 
filter transmission curves and directly comparing with the UVOT $u-v$ 
photometry to categorize the SNe as NUV-red or -blue. However, the 
exising UV spectral dataset has few spectra that span the 3000 -- 6000\AA\ 
wavelength range. 
As a less-favored alternative, we explore two methods;
comparisons of $u-b$ spectrophotometry with UVOT $u-b$ photometry (where
the separation between NUV-red and -blue is less) and truncating the UVOT
filter response at an intermediate wavelength to
generate $u-v(TR)$ spectrophotometry to compare with UVOT $u-v$ photometry.
A SN will be considered categorized if the two determinations agree.
The justification for having a truncated $v$ filter plus offset represent the 
$v$ filter is based on the findings of 
Foley et al. (2008), Cooke et al. (2010) and Maguire et al. (2012), who 
report relatively low scatter about a mean spectrum in that wavelength range.  

Foley, Filippenko \& Jha (2008) presented a complete collection of archival spectra 
obtained with the International Ultraviolet Explorer (IUE) 
or the instruments on HST up to 2004, yielding a sample of 20 SNe Ia. 
Ten spectra of 8 SNe~Ia (all HST spectra) 
had spectral coverage spanning 3000 -- 5500\AA\ at 
epochs earlier than +20 days, permitting $u-b$ and $u-v(TR)$ spectrophotometry 
during the epoch of interest. Foley et al. 2012a, 2012b and Foley \& Kirshner 2013 
presented UV spectra for SNe~2009ig, 2011iv and 2011by obtained with UVOT, HST and 
HST, respectively. 
The Palomar Transient Factory (PTF) is actively pursuing a program 
to discover SNe~Ia at very early epochs and to observe them in 
multiple wavelength ranges, including the UV. Cooke et al. (2010) presented 
HST-STIS spectra for 12 SNe Ia observed as part of this program, combining the near-peak 
spectra into a mean spectrum, which was compared with rest-frame 
NUV spectra obtained by ground-based telescopes from intermediate 
redshift SNe Ia. Each SN was observed with a single HST spectrum. 
All 12 SNe Ia had spectral
coverage spanning 3000 -- 5500\AA\, permitting $u-b$ and $u-v(TR)$ spectrophotometry.
UVOT photometry was obtained for 7 of the SNe Ia in that study, and 4 are 
included in this work (09dnl=09jb, 10icb, 10mwb=10gn, 10fps=10cr). 
 We include the narrow-peaked
SN~2010cr in these comparisons, but the color curves will studied in a separate paper
that concentrates on narrow-peaked SNe~Ia (Milne et al., in preparation).
Maguire et al. (2012) 
published a follow-up work with 16 additional SNe, and published peak widths and 
colors for all 28 SNe~Ia. We generate spectrophotometry from 25 of the 28 
SNe Ia in the Maguire et al. (2012) sample, excluding SNe~2010ju and  
PTF10zdk due to uncertainties in the peak dates, and SN~2010kg due to red $b-v$ 
colors in UVOT photometry.\footnote{There are indications that SN~2010kg might be a reddened 
MUV-blue event.}  

In order to have a sense of the level of agreement between spectrophotometry derived from
these spectra and UVOT photometry, we exploit a number of situations where UV spectra are
available for UVOT-observed SNe~Ia. We require the UVOT photometry to bracket the
spectral epoch. For 7 spectra of 7 SNe Ia from the Maguire et al. (2012) sample the weighted mean 
is consistent with no offset, -0.07 $\pm$ 0.10 mag, where the uncertainty is the standard deviation 
about the weighted mean. For 10 spectra of 3 SNe Ia from 
Foley et al. 2012a, 2012b and Foley \& Kirshner 2013, the weighted mean 
is again consistent with no offset, 0.04 $\pm$ 0.06 mag.  
Based on these comparisons we employ no offset between the datasets. 

The upper panel of Figure \ref{u_bv_cooke_foley} shows $u-b$ spectrophotometry for both the
Maguire and Foley archival samples, compared with UVOT $u-b$ photometry of the complete
collection of normal SNe Ia observed with UVOT. To determine NUV-blue versus -red for 
individual SNe, we calculate the
$\Delta (u-b)$ residual of each spectrophotometric point relative to a
linear fit to the UVOT photometry for NUV-red SNe Ia, offset by 0.1 mag to separate
NUV-red from NUV-blue (dashed line).
The lower panel of Figure \ref{u_bv_cooke_foley} shows 
$u-v(TR)$ spectrophotometry that truncates the UVOT filter reponse at 5500\AA\, 
compared with the full-filter UVOT $u-v$ photometry. An offset 
of 0.7 mag has been applied to match the two datasets, based upon the average flux lost 
by truncation for a sample spectrum that spans the $u$ and $v$ wavelength ranges. 
This correction is an $S$-correction. 
Once shifted, the distribution of 
HST spectrophotometry points follow the UVOT photometry. 
To determine NUV-blue versus -red for individual SNe, we calculate the $u-v(TR)$ 
residual of each spectrophotometric point relative to a 
linear fit to the UVOT photometry for NUV-red SNe Ia, offset by 0.2 mag to separate 
NUV-red from NUV-blue (dashed line). 
The color differences between the NUV-red and -blue events, shown in Figure \ref{hst_color_color},
lead to the NUV-red events being to the upper right 
and with positive residuals (Table \ref{specphot_resids}). 
From the residuals, 
we identify seven pairs of spectra that represent the two groups, but are
close in epoch. In Figure \ref{hst_compspec}, we show pairs of spectra that
are normalized in the 4000 -- 5500\AA\ wavelength range,
excepting the SN~1992A-09dnp
comparison, which was normalized between 4000 -- 4560\AA\. 
Clearly, the emission in the 3000 -- 3500\AA\ range leads to the
NUV-blue tendency, while the 3500 -- 4000\AA\ region is
complex. This suggests that photometry based upon the
3500 -- 4000\AA\ wavelength range would not detect a NUV excess.
The comparisons do not match optical light curve peak widths, as we have shown that to 
be of secondary importance for the UV-optical colors. 

Foley \& Kirshner (2013) present comparisons of SNe~2011fe and 2011by. We find both of 
these SNe to be NUV-blue events, with 2011fe bluer in the $uvm2-v$ colors. The 
spectral comparisons show SN~2011fe to have an excess for wavelengths shorter than 
2700\AA\, in general agreement with the UVOT photometry. These point to differences 
within the NUV-blue group. 
 
It underscores the effects of the dramatic color evolution
to note that the scatter of the $u-b$ photometry for
the HST observations made before +6 days would be 0.24 magnitudes if
the dramatic color evolution with epoch and NUV-blue membership were to
be ignored. However, the scatter of the two groups about two parallel
$u-b$ evolutions is only 0.08 magnitudes. Similarly, the scatter lowers
from 0.78 mag to 0.18 mag for the $u-v(TR)$ when epoch and
group membership are both accounted for. Studies that compare rest-frame
spectra obtained from SNe Ia at different redshifts must suitably
account for both the effects of color evolution with epoch and for
membership in the NUV-blue group.

\begin{figure*}[t]
\epsscale{1.8} \plotone{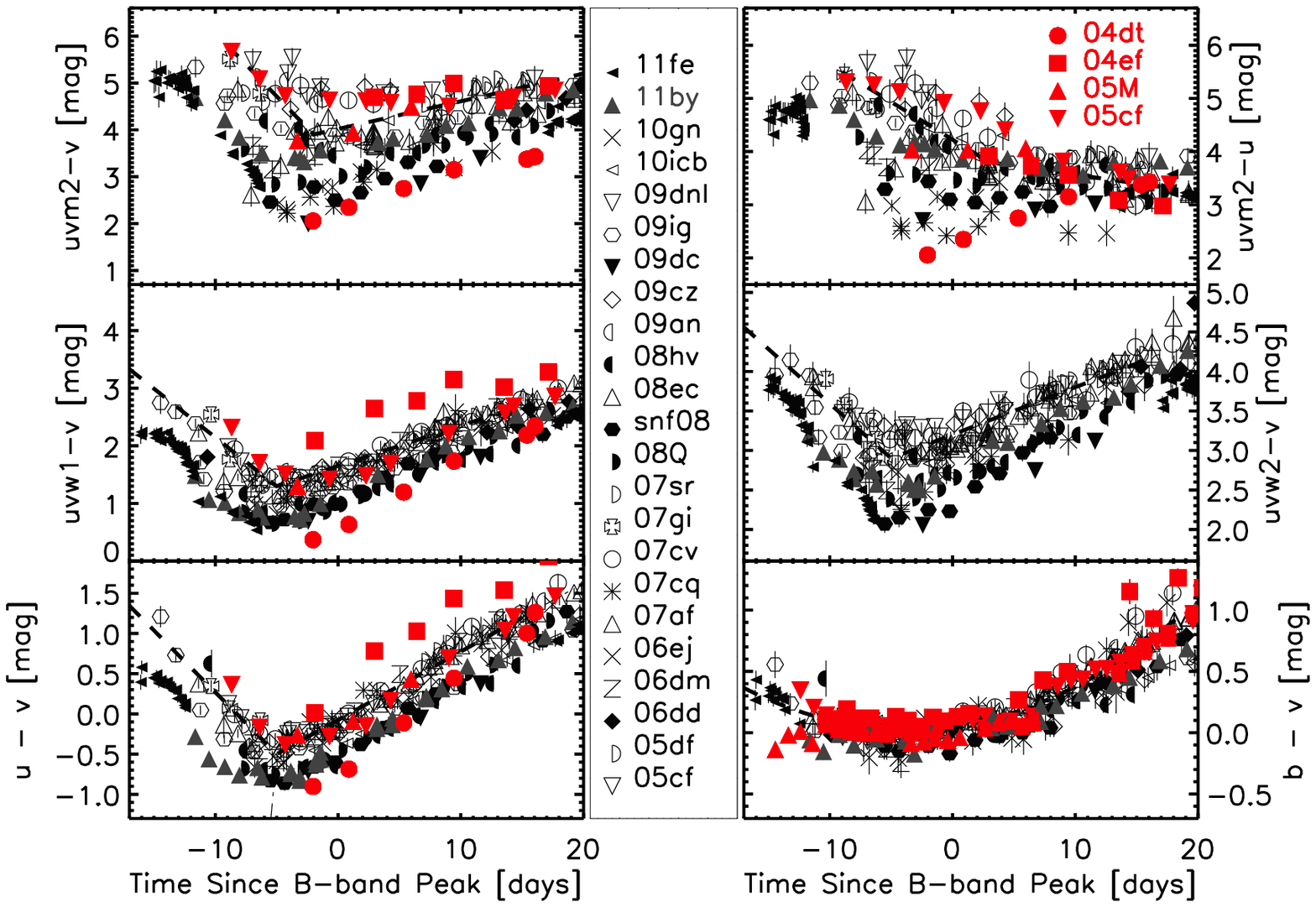}
\caption{Colors of 4 normal SNe~Ia from HST-ACS and KAIT photometry. The 
color curves from the HST/KAIT study are shown in red, color curves from 
UVOT photometry are shown in black. The HST F330W filter is compared the 
UVOT-$u$, the HST F250W to UVOT-$uvw1$ and HST F220W to UVOT-$uvm2$. 
SN~2004dt appears to be a NUV-blue event, 
while SNe~2005M and 2005cf appear as NUV-red events. SN~2004ef appears 
particularly red in the $u-v$ and $uvw1-v$ colors.}
\label{hst_acs_colors}
\end{figure*}

\begin{figure*}[t]
\epsscale{1.8} \plotone{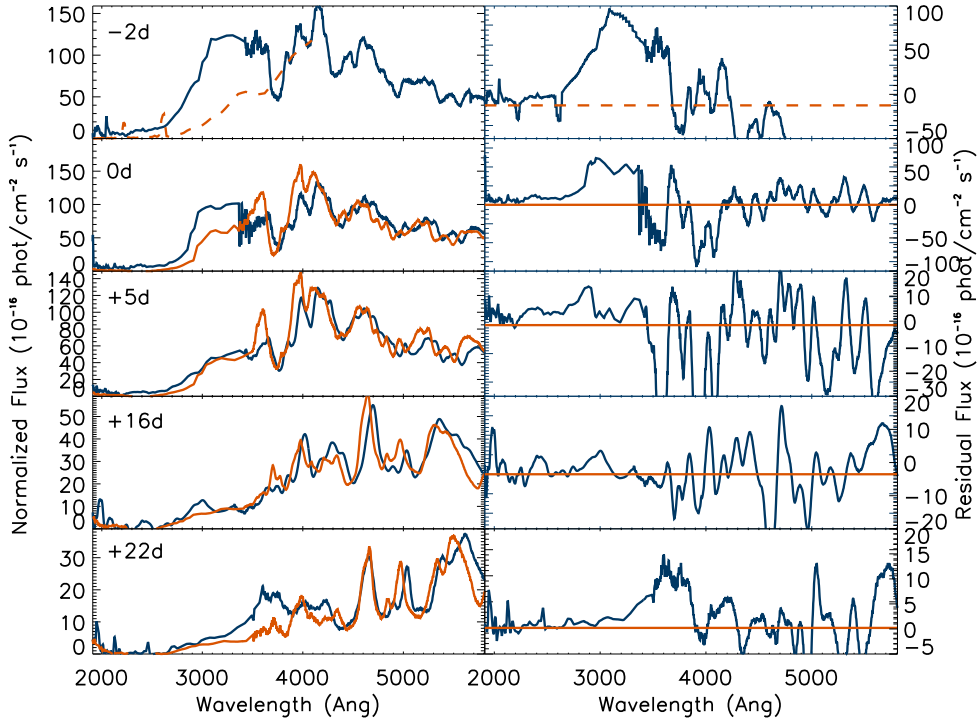}
\caption{Comparisons between HST-ACS spectra of the NUV-blue, SN~2004dt with 
two NUV-red SNe~2004ef and 2005cf. SN~2004dt spectra are shown as solid blue 
lines, SN~2005cf as solid orange lines and SN~2004ef as dashed orange lines. 
The left panels show the spectra normalized to emission in the 4000 -- 5000$\AA$ 
wavelength range and the right panels show 
the residuals of the NUV-blue spectrum relative to the NUV-red spectrum. For 
the early epochs, the SN~2004dt spectrum is brighter over a broad wavelength range. 
The SN~2004ef comparison was normalized at 4000 -- 4100$\AA$. }
\label{hst_acs_compspec}
\end{figure*}

It is interesting to revisit the work of Wang et al. (2012), who show 
HST-ACS grism spectra and HST UV photometry of 4 normal SNe Ia. 
Figure \ref{hst_acs_colors} shows colors of HST(UV) and 
KAIT(optical) photometry for the 4 SNe Ia in that study compared to the 
UVOT photometry. Despite making no effort to perform $S$-corrections between 
the F330W to $u$, F250W to $uvw1$ and F220W to $uvm2$ filters, the color 
evolution of 3 of the 4 HST SNe Ia follow the collective evolution of the UVOT sample. 
SN 2004dt is NUV-blue in that sample, while 2005M and 2005cf are 
NUV-red. SN~2004ef appears red in the $u-v$ and $uvw1-v$ colors, 
but fairly normal in the $b-v$ and $uvm2-v$ colors. The authors mention 
the UV faintness of SN~2004ef, and suggest a peak-width color relation as a 
possible cause. As will be shown in the next section, Figure \ref{color_vs_dmb15} 
shows that a peak-width color relation would not predict the red $u-v$ colors 
seen for SN~2004ef. Further, narrower-peaked SNe~Ia than SN~2004ef are featured in the 
UVOT sample. 
A spectral comparison of SN~2004dt versus two of the NUV-red SNe Ia supports 
an excess in the wavelength range shorter than 3500\AA\ as being responsible for the 
excess (Figure \ref{hst_acs_compspec} or Figure 5 from Wang et al. (2012)), 
at least at the earlier epochs. By 16 days after maximum light, the block of excess has disappeared, 
in agreement with Figure \ref{hst_acs_colors}, which shows the colors to be normal 
by that epoch. The dramatic reddening of SN~2004dt is different than the UVOT NUV-blue 
sample, which evolves largely parallel to the NUV-red events. 
Wang et al. (2012) interpret the NUV excess in the SN~2004dt spectra 
to be due to less iron-peak absorption 
in that wavelength range, relative to the NUV-red events, and emphasize that the 
spectral differences require further theoretical exploration.   
In the next section we will show that SN~2004dt differs from the UVOT NUV-blue sample 
based upon optical spectral features. 

SN~2011fe has been well observed with {\it Swift} and $HST$, suggesting that spectra for 
that NUV-blue event can be compared with other SNe~Ia well-studied by {\it Swift} 
(e.g. 2005cf: Bufano et al. 2009; 2009ig: Foley et al. 2012a; 2011iv: Foley et al. 2012b) 
and {\it HST} (e.g. 2011by: Foley \& Kirshner 2013). Spectral comparisons of SNe~Ia with 
well-sampled UVOT light curves, will avoid the need for truncated-filter color 
categorizations.

\section{Comparisons with Optical Parameters}

The existence of two groups of normal SNe Ia, separated by NUV-optical colors 
motivate the effort to correlate these events with parameters derived from 
optical light curves and spectra. We investigate the width of the optical light curves, 
blueshift of the SiII 6355\AA\ absorption line at maximum light and its time 
evolution, 
and the presence of CII 6580\AA\ absorption, as potential correlations. 

\subsection{First Parameter: Peak Width of Optical Light Curves}

\begin{figure*}[t]
\epsscale{1.0} \plotone{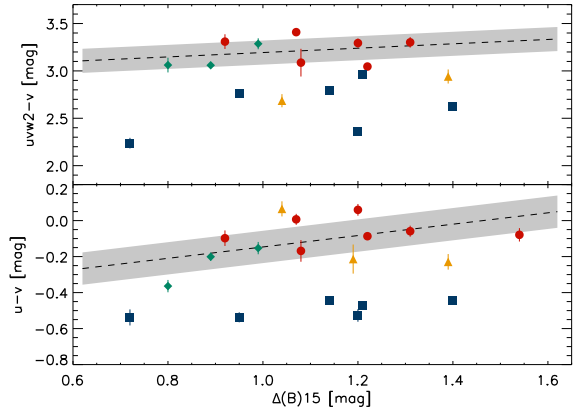}
\caption{$u-v$ colors at BPEAK plotted versus the optical peak width, 
$\Delta m_{15}(B)$. 
SNe~Ia are grouped as NUV-blue (blue-square),
NUV-red (red-circle), MUV-blue (orange-triangle) and irregular (green-diamond).
Both the NUV-red and NUV-blue groups appear to span the range of peak widths. 
A linear peak-width, peak-color relation for the NUV-red and irregular SNe~Ia 
is shown with 1$\sigma$ scatter. The linear relation is:  
(u-v)$_{BPEAK}$ = (0.33)$\Delta$ m$_{15}$(B) - 0.54.  }
\label{color_vs_dmb15}
\end{figure*}

The variation of the widths of the optical light curves, and the correlation 
of that width with the luminosity of the SN is a well studied 
subject. The reasonable starting point of a search for correlations between 
the UV-optical colors and optical emission is to compare the peak width of the 
B-band versus the $u$-$v$ color at BPEAK (as shown in 
Figure \ref{color_color_offset}). Figure \ref{color_vs_dmb15} 
shows the $uvw2-u$ and $u$-$v$ colors at BPEAK 
plotted versus $\Delta m_{15}(B)$. $uvw2-u$ was used rather than $uvm2-u$ due to 
the larger sample size. The NUV-blue group spans the majority of 
the range of peak widths of normal SNe Ia. The irregular subset of NUV-red 
events are the broadest-peaked events in that group, they combine to span the 
range of peak widths. The MUV-blue events are poorly sampled, but the two 
events with available $\Delta m_{15}(B)$ values appear towards the middle 
of the distribution. It is clear that the optical peak width is not 
correlated with the NUV-blue/NUV-red separation. The sense that the 
irregular minor group are the blue end of the NUV-red distribution 
suggests that the irregularities in the NUV-optical colors might just be a 
consequence of the larger $^{56}$Ni yield and higher explosion energies 
characteristic of the broad-peaked side of SN~Ia light curves. 

\subsection{Second Parameter: SiII velocity vs NUV-optical color}

Years before the reports of unburned carbon as a ``second parameter" 
of SNe Ia, independent of the LWR, Benetti et al. (2005) recognized 
the phase evolution of the blueshift of the SiII $\lambda$6355\AA\ 
line as a potential second parameter. Benetti et al. (2005) separated  
normal SNe Ia based upon $\dot{v}$, the rate at which the blueshift 
decreased with epoch. The cut between high-velocity gradient (HVG) and
low-velocity gradient (LVG) events was set at 80 km/s/day. 

With the goal of minimizing the number of spectra required to classify a 
SN according to the Benetti et al. (2005) second parameter, Wang et al. (2009b) 
developed the blueshift of the SiII $\lambda$6355\AA\ near-peak as a proxy 
for the velocity gradient. HVG events also feature highly blueshifted 
SiII at peak, so that HVG=HV (High Velocity), while LVG=NV (Normal Velocity). 
Wang et al. (2009b), Foley, Sanders \& Kirshner (2011)  
and Foley \& Kasen (2011) investigated the impact of this second parameter 
on SNe~Ia as distance indicators, finding HVG/HV events redder at peak 
reduces the Hubble residuals when treating the two groups 
independently.    

In columns 5-7 of Table \ref{opt_params}, we list the available 
HVG/LVG (5) and HV/NV (6) determinations for UVOT SNe~Ia. It is interesting that 
many of the NUV-blue events are LVG/NV while none are HVG/HV. 
The NUV-red events show equal numbers of LVG/NV versus HVG/HV events, 
a trend also present in the poorly sampled irregular and MUV-blue 
groups. The right panel of Figure \ref{bar_graph} graphically shows the 
search for a correlation. 
Collectively, this suggests that the velocity gradient is not a reliable 
predictor of NUV-optical colors, but that NUV-blue events might be a subset of 
the LVG/NV group. 

Although none of the UVOT NUV-blue sample are HVG events, 
SN~2004dt, shown in Section 3 to be a NUV-blue event, is a HVG event.  

\subsection{Second Parameter: Unburned Carbon vs NUV-optical Color}

Four recent papers have searched for the CII 6580\AA\ absorption feature in 
optical spectra. Despite the complications presented by the nearby, strong 
SiII 6355\AA\ absorption line, Parrent et al. (2011), Thomas et al. (2011), 
Folatelli et al. (2012) and Silverman \& Filippenko (2012) have all 
reported detection of that line in 20\%-35\% of the events for which 
early-phase spectra are available. 
Parrent et al. (2011) searched archival spectra, 
Thomas et al. (2011) searched Nearby Supernova Factory (NSF) spectra, 
Folatelli et al. (2012) searched Carnegie Supernova Project (CSP) spectra,  
and Silverman \& Filippenko (2012) searched Berkeley Supernova Ia 
Program (BSNIP) spectra. 
Estimation of what fraction of SNe Ia feature the CII 6580\AA\ is complicated 
by observational biases due to the line fading with epoch (disappearing by 
roughly optical maximum in most cases), failure to 
extract the line in the presence of high-velocity SiII line absorption and 
often low S/N of the spectra obtained, which might explain the range of 
fractions. Further, 
a unified methodology has not yet been established to quantify the 
strength of the CII 6580\AA\ line at a standard epoch. 
Parrent et al. (2011) and Thomas et al. (2011) simply listed 
whether the line was detected, while Folatelli et al. (2012) 
introduced an intermediate situation, where the CII feature leads to 
only a flat edge to the SiII 6355\AA\ line, rather than an absorption
line at 6580\AA\ .  Silverman \& Filippenko (2012), follow a similar methodology 
as Folatelli et al. (2012).\footnote{Folatelli et al. (2012) 
estimate a pseudo equivalent width (pW), and report the evolution with 
phase of pW, for SNe Ia with multi-epoch spectroscopy. Future campaigns  
will attempt to obtain the time evolution of pW for UVOT SNe Ia.} 
Collectively, the studies demonstrate that the presence or absence of 
unburned carbon features are a distiguishing characteristic in optical 
spectra of SNe Ia, but a characteristic that is observationally very 
difficult to quantify. 

The first suggestion that there might be a correlation between unburned 
carbon and NUV-blue SNe Ia was made by Thomas et al. (2011), who 
reported that SNe~2008Q, 2008hv, SNF2008-0514-002, and 
2009dc, all UVOT NUV-blue events, feature CII absorption. SN~2011fe has 
also been reported to have CII absorption (Nugent et al. 2011), as 
has SNe~2006dd \& 2011by,  meaning 
that all 7 UVOT NUV-blue SNe Ia have CII absorption. Further, SN~2001ay, 
the NUV-blue event from the HST sample of Foley et el. (2012c) features 
CII (Krisciunas et al. 2011), as does PTF09dnp (Parrent, private 
communication). SN~2004dt is the only NUV-blue event without a CII 6580\AA\ 
detection, with both the Parrent et al. (2011) and Folatelli 
et al. (2012) studies reporting a non-detection. Combined with it is being the only 
HVG event that has been found to be NUV-blue, it defies both correlations seen 
in the UVOT sample. However, Altavilla et. al. 
(2007) reported detecting oxygen absorption lines. As oxygen is an 
ambiguous element, because it could be due to unburned ejecta or due to 
partial burning, SN~2004dt remains a puzzle. SN~2004dt defies a peak 
width, SiII polarization correlation relation (Patat et al. 2012), so 
perhaps it is more of a peculiar event than we have recognized it to be. 

\clearpage

\begin{figure*}[t]
\epsscale{1.8} \plotone{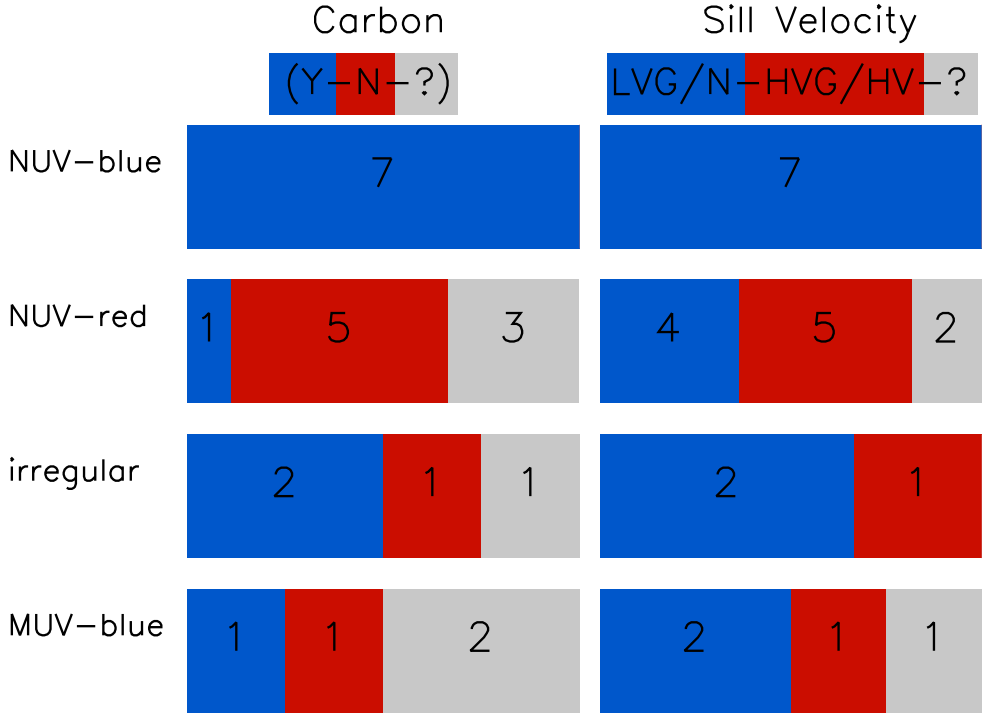}
\caption{Comparisons of UV-optical colors with optical parameters for 
21 SNe~Ia in the UVOT sample. 
Detections of unburned carbon (Y, A, F or P in Table \ref{opt_params}) 
and LVG/N SiII velocity gradients are shown in blue, while non-detections of 
unburned carbon (N in Table \ref{opt_params}) and HVG/HV SiII velocity gradients 
are shown in red. Undetermined cases are shown in grey. The numbers of events are 
shown in each region. NUV-blue events tend to have unburned carbon, while NUV-red 
events tend towards non-detections. NUV-blue events tend to be LVG/N events, 
but NUV-red events are equally likely to be LVG/N or HVG/HV.}
\label{bar_graph}
\end{figure*}

\begin{figure*}[t]
\epsscale{1.8} \plotone{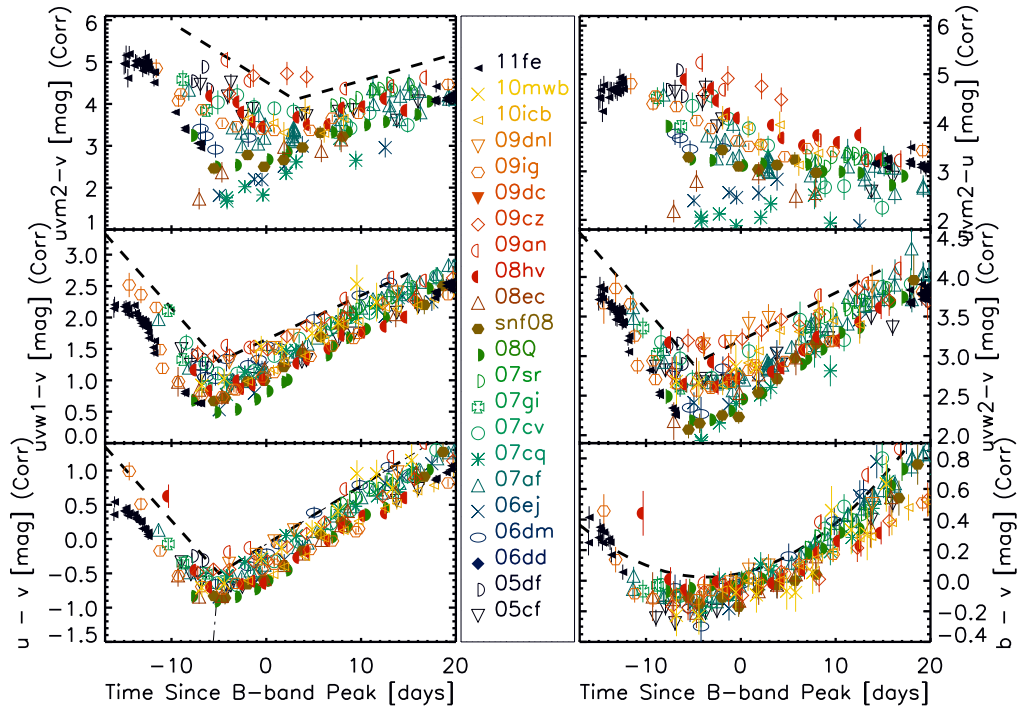}
\caption{Near-peak colors of seventeen normal SNe~Ia compared to the 
$v$ band with reddening correction applied. The symbols are the same as Figure 
\ref{col_N_zoom}. The individual color curves show larger variations than seen 
in Figure \ref{col_N_zoom}.}
\label{corr_N_zoom}
\end{figure*}

\begin{figure*}[t]
\epsscale{1.8} \plotone{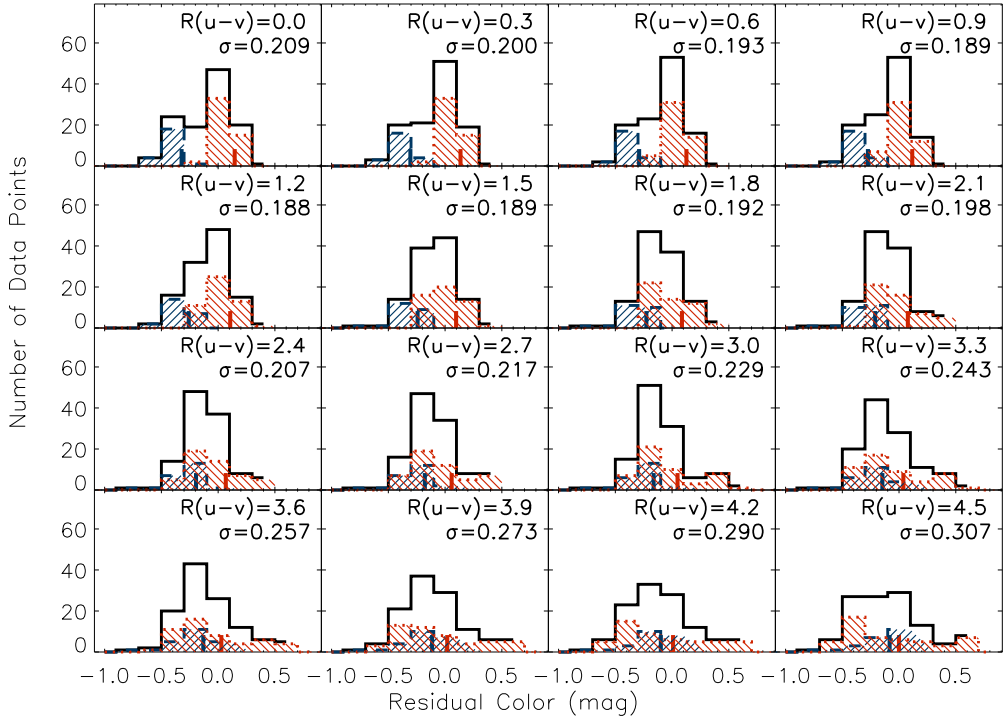}
\caption{Histograms of residuals of peak-colors of SNe Ia relative to a linear function 
from -5 -- +10 days for different choices of R$_{u}$-R$_{v}$. The slope of the color 
evolution is fixed, with the offset being fitted for each choice of R$_{u}$-R$_{v}$. The 
blue-dashed histograms show data for NUV-blue SNe, the red-dashed histograms show 
data for NUV-red SNe. The black-solid histogram is the combined data. Blue and red ticks 
show the mean residuals for NUV-blue and -red data, respectively. The scatter for the total 
data is shown for each choice of R$_{u}$-R$_{v}$.}
\label{dustlaw_u}
\end{figure*}

\begin{figure*}[t]
\epsscale{1.8} \plotone{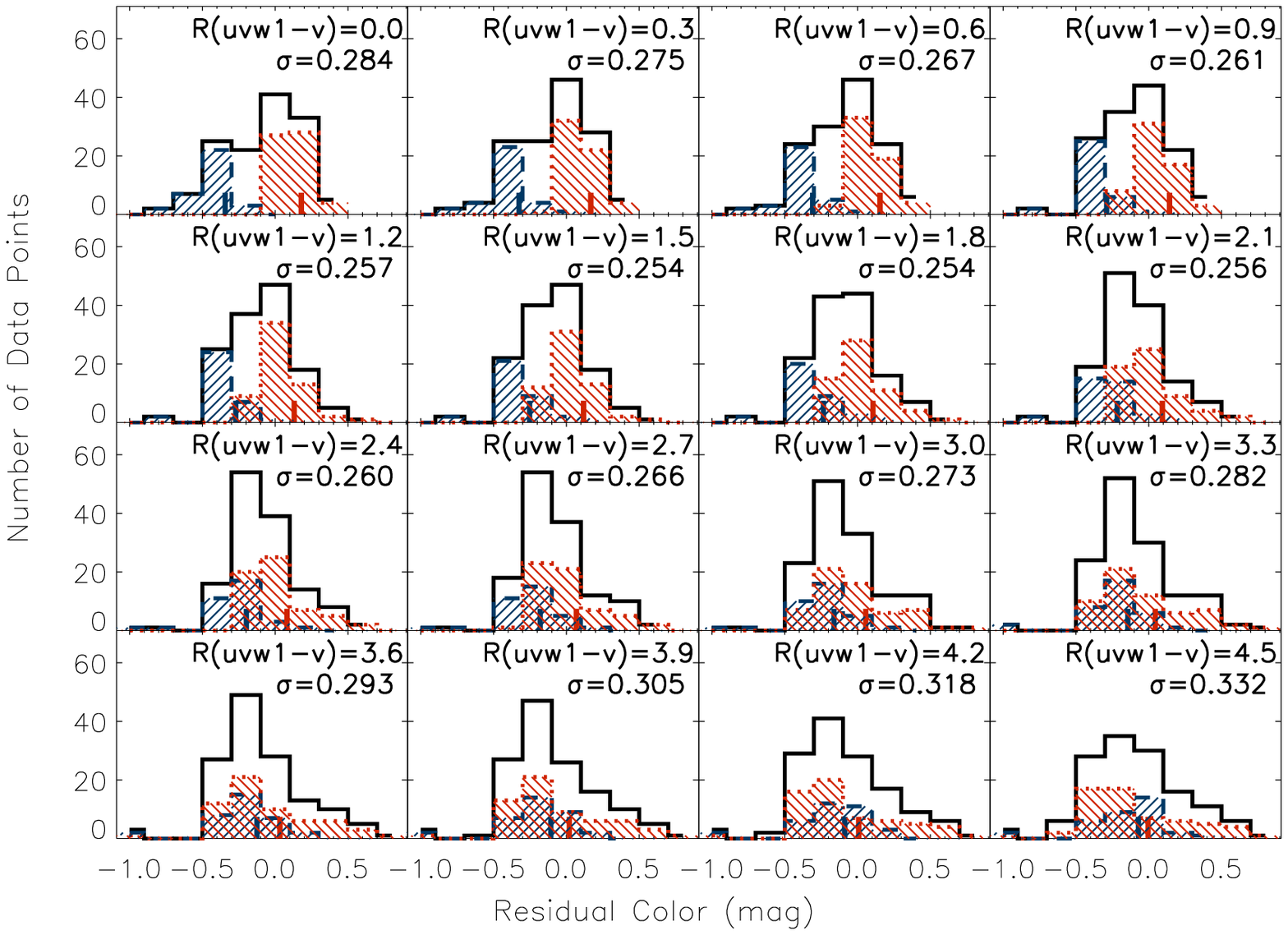}
\caption{Same as Figure \ref{dustlaw_u} for different choices of R$_{uvw1}$-R$_{v}$}
\label{dustlaw_uvw1}
\end{figure*}

Demonstrating the inverse, that all NUV-red events lack unburned carbon is 
more complicated, as the CII feature fades with phase, and is contaminated 
by the SiII 6355\AA\ line. A complete treatment of the possibilities is 
beyond the scope of this work, but we show the reported outcomes of 
searching for this feature in the studies mentioned above, in Table 
\ref{opt_params}: columns 3 \& 4 and the left panel of Figure \ref{bar_graph}. 
The majority of NUV-red SNe~Ia have 
non-detections of CII, but SN~2005cf is NUV-red with a CII detection 
reported by two groups. The irregular and MUV-blue groups are too 
poorly sampled to draw conclusions beyond the fact that both groups 
feature a detection and a non-detection of the CII line. 
After explaining the sharp evolution of UV-optical colors, 
theoretical modelling of SN Ia emission is presented with a second 
challenge: demonstrating why NUV-blue events are very likely to 
feature unburned carbon while NUV-red are unlikely to feature 
unburned carbon. 

The path forward in this search for correlation is to continue to 
obtain UVOT photometry for nearby SNe Ia found at the early epochs 
for which CII identification is possible. As the NUV-blue color persists 
during the entire peak epoch, if a strong correlation were to be established, 
the UV-optical colors could be used to identify the SN Ia explosions of 
this variety. It is tantalizing that this study finds that roughly one-third 
of normal SNe Ia are NUV-blue, which is in the range of events reported to feature 
unburned carbon. There is no clear reason why the sub-samples for which the 
Si velocity blueshift categorization and/or the existence/absence of unburned carbon 
are known should differ from the larger UVOT sample, so we suggest that the 
tendencies discussed are representative of SNe~Ia as a whole. The ratio of LVG/NV 
to HVG/HV SNe~Ia is roughly 2:1, in agreement with the findings of Wang et al. (2009b).

\subsection{Other spectral or light curve features}

The definition of ``normal" used in this paper is very permitting, by eliminating 
only narrow-peaked and SN~2002cx-like events. There are other differences in the 
photometry and spectra of this normal class that warrant some discussion. 
The MUV-blue SN~2007cq was considered SN~1991T-like due to the early appearance of absorption 
features from iron-peak elements, while exhibiting less SiII absorption (Blondin et al.2012). 
However, there is no information for SNe~2006ej or PTF10mwb that suggests those SNe~Ia  
were similar events, indeed Silverman, Kong \& Filippenko (2011) classify SN~2006ej as 
normal and not 91T-like. 

SN2001ay was reported 
by Krisciunas et al. (2011) to be the ``most slowly declining type Ia supernova" and 
thus an extreme of peak-width. However, the absolute optical magnitudes were in the 
normal range, making SN~2001ay a faint outlier in the LWR. 
SN~2009dc is considered a candidate for a super-Chandrasekhar 
explosion based on the high-luminosity and slow declining light curve (Taubenberger et al. 2010; 
Silverman et al. 2011). Both SNe~Ia are NUV-blue, but have color evolution that does not 
differ appreciably from other NUV-blue events with much narrower peak-widths. 
In a separate work, Brown et al., in preparation, will look at the UVOT sample of 
super-Chandrasekhar candidates, finding them also blue in NUV-optical colors, but 
exploring whether as a group they are too blue to categorize them as part of this 
NUV-blue group. 

SN~2009ig was reported to feature an apparently broad SiII absorption line at very early epochs 
(Foley et al. 2012a). 
This could be due to two regions of SiII absorption at different velocities or due to non-SiII 
contamination of the line. A flattened SiII absorption feature was also reported for SN~2009cz 
(Kankare \& Mattila 2009, in CBET 1763). Both SNe~Ia are NUV-red-irregular, but information 
is lacking on whether the other two members of that minor group have that feature. 
Foley et al. 2012 argued that the lack of very early epoch spectra allow that broad SiII 
absorption might be a common feature in SNe~Ia, and thus not a distinguishing characteristic 
of a minor group. However, Mazzali et al. (2005) find that high-velocity SiII features in 
early spectra are not common.  

None of these comparisons have yielded strong correlations, but as the optical and UV datasets 
increase, potential correlations should become more apparent. H\H{o}flich et al. (2010) 
showed differences in the light curve shapes between SNe~Ia with similar stretch values. In 
particular, different shapes of the rise to peak were seen between events with identical 
decline rates. 
As this study has concentrated on colors rather than light curve fitting, we will address 
this in a future work that concentrates more on fitting light curves.

\section{Extinction} 

The distinguishing feature between NUV-blue and NUV-red SNe Ia is primarily an 
offset in NUV-optical colors. This is a potential complication, since the 
observed colors from the SN~Ia sample would be altered by the presence of 
dust along the line of sight. Attempting to minimize this complication, in this 
work we select only SNe~Ia with estimated reddening of E(B-V) = 0.25 mag, or less. 
The selective extinction, R$_{i}$, for the 
UVOT filters was estimated in B10 for two assumptions of the nature of the UV extinction 
law, and large values of R$_{i}$ were derived for each assumption. 
This translates to the UV wavelength range 
being highly sensitive to extinction and places importance on accurate reddening estimation.
Determination of host galaxy reddening largely relies upon relations that compare 
observed optical colors of a given SN with the colors of supposedly zero-extinction SNe~Ia. 
This is performed at optical peak for the $B-V$ Phillips relation (Phillips 1999), 
between 30 and 90 days after maximum light for the $B-V$ Lira-Phillips tail relation (Lira 1995), 
or for the entire $UBVRI$ light curve out to +90 days for MLCS2k2 (Jha, Riess \& Kirshner 2007). 
These techniques are employed to produce the E(B-V) estimates shown in Table 1. 
We gave high priority to published extinction estimates, followed by MLCS2k2(R$_{V}$=1.7) 
values provided to us (see B10). When no other estimates were available, we used UVOT 
$b$ and $v$ photometry to employ the Phillips peak or Lira tail relations. 
These estimates emphasize the best information rather than attempting to be homogeneous in 
method. 

Until recently, normal SNe~Ia have been treated as one group, with no distinction made between 
potential second parameters, such as LVG/HVG or the presence/absence of unburned carbon. 
A number of recent studies have drawn into question both the treatment of normal SNe Ia as 
a single group, citing differences in the peak colors and favored extinction law. Although 
our sample is inadequate to definitively study peak optical colors and extinction as a function of 
NUV-optical colors, it is important that we address extinction as a function of sub-groups 
within the normal class, to the extent that it can be addressed at the current time. 

\subsection{Preliminary Extinction Correction}

Accepting for the moment the practice of treating all normal SNe~Ia by a single reddening law,
in Figure \ref{corr_N_zoom} we show the color curves of the SNe~Ia in our sample with
E(B-V)$\leq$0.5 mag. The colors have been corrected according to the CSLMC dust law using mean
values of R$_{i}$-R$_{v}$ from the normal SNe Ia in Table 6 of B10: 
5.0,9.5,3.8,2.9,1.3 for $uvw2-v$,$uvm2-v$,$uvw1-v$,$u-v$,$b-v$,
respectively. The color differences between NUV-red and NUV-blue SNe
are much less convincing for Figure \ref{corr_N_zoom} than for Figure \ref{col_N_zoom}. 
The fact that a considerable portion of the difference between the two groups goes away 
requires us to address the suggestion that the NUV-red and -blue groupings could be an 
illusion generated by failing to properly correct for reddening.
One way to address that suggestion is to ignore the specific E(B-V) estimates and study the 
effect of reddening on color-color plots.  
Figure \ref{color_color_offset} shows reddening vectors for the MWG dust law (black-solid) and 
CSLMC dust law (red-dashed) emanating from the 
colors of SN~2011fe. For a few color pairs, the vectors connect NUV-blue and -red, but across 
all color pairs, neither dust law produces vectors that would reconcile the two groups.

\subsection{Varying Extinction Law rather than NUV-red/-blue}

The failure of the reddening vectors to line up with NUV-red events could simply point to the 
utilization of incorrect selective extinction values. The ongoing debate as to whether the 
optical light curves of SNe~Ia require selective extinction values significantly different 
than found for the Milky Way certainly allows for that possibility. To address that possibility, we 
fit the $u-v$ colors of the entire -5 day to +10 day dataset with a linear relation,  
and study the distribution of the residuals of each group as a function of the selective extinction 
value, using the E(B-V) values from Table 1.  
If the NUV-red/-blue grouping is an illusion and a single choice of R$_{u}$-R$_{v}$ for all 
SNe~Ia is appropriate, the NUV-red and NUV-blue distributions will match with the correct choice 
of R$_{u}$-R$_{v}$. 
In Figure \ref{dustlaw_u}, R$_{u}$-R$_{v}$ is increased from 0.0 (no correction) to 4.5, with 
separate histograms for NUV-red, NUV-blue and the combination of the two. 
The choice that leads to the lowest scatter ($\sigma$=0.19 mag), R$_{u}$-R$_{v}$ = 1.2, still 
features NUV-blue events as bluer than NUV-red events. For reference, the scatter that results from
treating NUV-red and -blue as different groups is lower, only 0.08 mag. 
The NUV-red and -blue distributions have the same peak for 2.4 $\leq$ R$_{u}$-R$_{v}$ $\leq$ 3.6, 
(which contains the CSLMC dust law value), but the scatter is quite a bit larger than the two-group 
result or for the lowest scatter one-group result. 
The same tendencies are seen for R$_{uvw1}$-R$_{v}$ (Figure \ref{dustlaw_uvw1}). 

Explaining the 
observed color differences as due purely to host galaxy reddening would seem to require 
multiple reddening laws, rather than a single reddening law. If the  
differences between the NUV-red and -blue groups were to only be present as a color offset, 
it would be plausible to accept that Figure \ref{col_N_zoom} is an illusion created by 
difficulties determining E(B-V) and R$_{i}$-R$_{v}$ for each SN~Ia in the sample. However, 
Sections 3 \& 4 have established other differences between the groups that argue for the 
reality of the groups. We now explore the possibility that the extinction estimates are 
flawed by the failure to recognize these differences. 

\subsection{Alternative Extinction Estimation Methods}

The problem with the methods mentioned earlier in this section is that the above 
relations treat all normal SNe~Ia as 
varying only by peak width. If there are $B-V$ color differences between LVG and HVG SNe~Ia, 
as argued by Wang et al. (2009b,2013) the above relations in their current form do 
not distinguish. If there are $B-V$ color differences between SNe~Ia with unburned carbon and those 
without, as argued by Folatelli et al. (2012) the above relations in their current form do
not distinguish. Most importantly, if there are $B-V$ color differences between NUV-blue and NUV-red 
SNe~Ia, as suggested by Figure \ref{color_color_offset}, the above relations in their 
current form do not distinguish. A systematic error in reddening estimates for 
one of the NUV-red/-blue groups, would be a natural explanation for why reddening-corrections lead 
to less order in the color curves. 

Support for color differences between NUV-red/-blue events can be seen in Table 1, where 
the mean host galaxy extinction for NUV-blue events is estimated to be only -0.05 mag, which is  
0.15 mag less than the mean extinction estimated for the other SNe~Ia. 
It is possible that there is less host galaxy dust for NUV-blue events, but it seems more 
likely that these events are intrinsically bluer than NUV-red events in B-V leading to underestimation 
of the host galaxy reddening for NUV-blue SNe~Ia. Doubling or tripling that E(B-V) difference 
via the R$_{i}$ terms might be the reason that extinction correcting appears to negate the 
existence of separate NUV-optical color groups. 

Simply drawing into question the extinction estimates for the NUV-blue group might solidfy 
the argument for the existence of different groups, but it does not address determination 
of improved E(B-V) estimates. Accurate extinction estimation is critical for exploring 
the absolute magnitudes of NUV-blue versus NUV-red SNe, so further investigation is 
important.   
One way to explore this issue is to rely on the correlations found in the previous section, 
in an attempt to improve the extinction estimates for the UVOT sample by utilizing 
larger published studies. Folatelli et al. (2012) found that the $B-V$ peak pseudo-colors of 
SNe~Ia with unburned carbon are bluer than those without unburned carbon. The strong 
correlation we find between unburned carbon SNe and NUV-blue SNe is thus support for the 
idea that NUV-blue events have bluer $B-V$ peak colors. However, Folatelli also concentrated 
on a ``low-reddening" sample, where the SN location in the host galaxy and the the lack of 
Na I D absorption was suggestive of minimal reddening. Dividing the low-reddening sample into 
with and without CII sub-samples, intrinsic peak-width versus peak pseudo-color relations were 
derived. The sampling is small, with six events per sub-sample, but the primary difference 
between the groups is the dependence on the peak width, rather than an offset between the 
groups. 

Wang et al. (2009b, 2013) studied the peak colors of their large sample of SN~Ia light curves, separated 
by the HV/NV determinations. They found that HV events were redder in peak B-V by ~0.1 mag than the 
larger NV group. Since HV/NV is a proxy for HVG/LVG, and we have shown that UVOT NUV-blue events 
are LVG rather than HVG, it is possible that NUV-blue events are the cause of the LVG/NV group 
being bluer. An argument against that interpretation is found in the work of Foley \& Kasen (2011), 
who found lower Hubble residuals in the NV group than for the HV group for a range of 
B$_{max}$-V$_{max}$ colors, although the NV group contains both NUV-blue and NUV-red events, 
while the HV group contains only NUV-red events. The HV group would have to possess much larger 
intrinsic scatter to exceed the scatter of the NV group that would emcompass two subgroups with 
different peak colors.   

The discussions in this section do not establish the reality of systematic extinction 
underestimation errors for NUV-blue SNe Ia. This means that the existence of the NUV-blue and 
NUV-red groupings cannot be stated as having been proven. They do however, suggest that extinction 
estimation errors are a plausible explanation for the failure of color corrections 
to reduce scatter in the UV-optical color curves, relying upon the UV spectral differences as 
an independent proof of the existence of two groups. 
One idea worthy of further investigation is that
the NUV-optical colors reveal a homogeneity within each of the two groups of normal
events that can be exploited to {\bf determine} the extinction. This inversion
of the problem should be explored with rest-frame UV observations of many more
SNe~Ia.

\section{Discussion}

Motivated by the desire to advance the cosmological use of SNe Ia, 
emission in the rest-frame wavelength ranges blueward of the $B$-band 
has recently been studied to 
determine the level of scatter within normal SNe Ia about mean spectra and 
light curves. The consensus has been that emission blueward of the $B$-band
features considerable scatter, making that wavelength range less promising 
for standard candle cosmological applications. Based upon thousands of observations with 
the {\it Swift} UVOT instrument, we report that much of the scatter in the 
$u$ and NUV wavelength ranges results from the existence of two groupings of normal 
SNe Ia, based upon the NUV-optical colors. Combined with the dramatic 
time-evolution of the NUV-optical colors for both groups, we suggest that 
the NUV wavelength range deserves a closer look as to whether it is 
cosmologically useful.   

UVOT photometry of SNe Ia has established that the NUV-optical colors change 
dramatically with epoch, initially becoming blue and transitioning quickly to 
redder colors. This mimics the same trend seen in $B-V$ colors, but is more 
pronounced.  The addition of UVOT photometry of recent SNe Ia has permitted 
the determination that there exist two different major groupings of NUV-optical 
colors for normal SNe Ia, groups we name ``NUV-blue" and ``NUV-red". The color 
differences extend over multiple NUV filters and both groups are observed across 
the entire range of redshifts accessible to UVOT, suggesting that $K$-corrections 
are not the cause. 
There is some evidence for two minor groups, one ``MUV-blue" group that is blue 
only in the bluest $uvm2$ and $uvw2$ filters, and one ``irregular" group that 
appears as something of a transitional group between the NUV-blue and NUV-red 
groups. In a number of figures, the irregular appeared as an extension of the 
NUV-red group, suggesting that it is likely a consequence of higher $^{56}$Ni 
yields for the broadest-peaked NUV-red events.     

We have searched suggested second parameters for a correlation with the 
NUV groups. While all UVOT NUV-blue events are of the low-velocity (LVG/N) SiII 
blueshift group, we find that many of the NUV-red events are also of the 
low-velocity SiII group. NUV-blue events appear to be a subset of the 
low-velocity group. A correlation is found with the detection of 
unburned carbon (CII) optical absorption lines, where all NUV-blue and 
MUV-blue UVOT SNe~Ia feature that absorption line, but there is only 
a single suggestion of that feature in the larger NUV-red group. 
This supports iand expands upon the findings of Thomas et al. (2011), who first 
noted the correlation.

Creating spectrophotometry from HST UV spectra of SNe~Ia, we have 
determined that 8-12 events are of the NUV-blue 
group. Comparing spectra of those SNe with NUV-red events at a similar 
epoch, normalized on the optical emission, we find that the NUV excess is 
generated by a block of excess from 2900--3500\AA\, in agreement with 
the findings of Wang et al. (2012) for the NUV-blue, SN~2004dt. As emission 
in that wavelength range is strongly affected by iron-peak element absorption, 
the NUV excess might point to less fully synthesized material near the 
surface of the ejecta for NUV-blue events. When combined with the apparent 
correlation with unburned carbon, this might point to the burning front 
not reaching as near the surface as in NUV-red events (i.e. governed by 
the explosion physics), or to the presence of additional unburned carbon outside 
of the explosive burning (i.e. a signature of a double-degenerate explosion). 
The distinct separation of the two groups would seem to suggest against viewing 
angle as the cause, but this needs to be explored further. There is little 
spectral information as to the nature of the difference between NUV-blue and 
MUV-blue SNe.

The dramatic color evolution and existence of two groups of NUV-optical 
colors within the class of normal SNe~Ia is important for the creation 
of mean UV spectra of normal SNe Ia, and attempts to utilize mean spectra 
to search for variations of the UV emission within SNe~Ia of a given 
redshift. The fairly high homogeneity within each NUV-optical color group 
gives rise to optimism that with more work, the NUV wavelength range might 
be an important contributor to SN~Ia cosmology.  

SN Ia observing continues with UVOT. The first 3 years of the 
UVOT SN Ia survey featured a sample similar
in number of targets and number of observations as the optical SN Ia
surveys of the late 1990s (Hamuy et al. 1996a, 1996a, 1996bb, Riess et al. 1996).
Combined with ongoing 
HST NUV campaigns on nearby SNe~Ia, significant progress should be made 
towards better understanding SN~Ia progenitor systems and the nature of 
a SN~Ia explosion.

\bigskip
P.A.M. acknowledges support from NASA ADAP grant NNX10AD58G. 
F.B. acknowledges support from FONDECYT through Postdoctoral grant 2130227.
P.A.M thanks R. Foley, K. Maguire, J. Cooke, X, Wang, J. Silverman, M. Ganeshlingham, 
R.C. Thomas, J. Parrent and H. Marion for assistance accessing spectral and 
photometric datasets critical for characterizing each supernova. P.A.M. thanks 
P. Mazzali for text relating to line-blocking and line-blanketing effects, 
and R. Foley \& J. Silverman for discussions of interpretation of non-UVOT 
datasets. All 
supernova observers thank the mission operations team at Penn State for 
scheduling the thousands of individual UVOT target-of-opportunity observations 
that comprise this dataset. The NASA/IPAC Extragalactic Database (NED) was 
utilized in this work. NED is operated by the Jet Propulsion Laboratory of the 
California Institute of Technology, under contract with the National 
Aeronautics and Space Administration.

\begin{table*}
\scriptsize
\vspace{-9mm}
\caption{NUV-Optical Colors of HST Spectral Sample}
\begin{tabular}{l|c|cc|cc}
\hline
\hline
SN & t(BPEAK) & $u-b$ & $\Delta[u-b]^{a}$ & $u-v$(TR5500) & $\Delta[u-v(TR5500)]^{b}$ \\
Name & [days] & [mag] &  [mag] &  [mag] &  [mag] \\
\hline
\multicolumn{6}{c}{NUV-Blue}  \\
\hline
PTF09fox &  2.60 & -0.24 & -0.07 & -0.42 & -0.34 \\
PTF10ufj &  2.70 & -0.05 &  0.12 & -0.94 & -0.17 \\
PTF09dlc &  2.80 & -0.25 & -0.09 & -0.35 & -0.28 \\
PTF10hdv &  3.30 & -0.20 & -0.07 & -0.13 & -0.11 \\
PTF10qjq &  3.50 & -0.36 & -0.24 & -0.05 & -0.05 \\
PTF10wnm &  4.10 & -0.31 & -0.22 & -0.33 & -0.38 \\
PTF09dnp &  5.80 & -0.37 & -0.37 & -0.12 & -0.31 \\
PTF10ndc &  5.80 & -0.20 & -0.20 & -0.50 & -0.69 \\
\hline
\multicolumn{6}{c}{NUV-Blue or Irregular$^{c}$}  \\
\hline
PTF10qjl &  5.90 & -0.11 & -0.11 &  0.03 & -0.17 \\
PTF10qyx &  6.80 & -0.05 & -0.10 & -0.05 & -0.33 \\
SN2001ay &  8.30 & -0.20 & -0.34 &  0.34 & -0.07 \\
SN2001ay & 15.00 &  0.02 & -0.48 &  0.74 & -0.24 \\
\hline
\multicolumn{6}{c}{NUV-Red}  \\
\hline
PTF10mwb$^{d}$ & -0.40 & -0.34 & 0.01 & -0.81 &  0.23 \\
PTF09dnl &  1.30 & -0.23 &  0.01 & -0.70 &  0.19 \\
PTF10hmv &  2.50 & -0.10 &  0.07 & -0.46 &  0.33 \\
PTF09foz &  2.80 & -0.07 &  0.09 & -0.63 &  0.13 \\
PTF10xyt &  3.20 & -0.02 &  0.12 & -0.32 &  0.41 \\
PTF10tce &  3.50 & -0.04 &  0.09 & -0.43 &  0.27 \\
SN2001ba &  4.00 & -0.04 &  0.06 & -0.45 &  0.21 \\
PTF10wof &  5.90 &  0.23 &  0.22 & -0.11 &  0.39 \\
PTF10nlg &  6.20 &  0.41 &  0.39 &  0.06 &  0.54 \\
PTF10yux &  7.10 &  0.36 &  0.28 &  0.17 &  0.57 \\
SN2001eh &  8.10 &  0.18 &  0.06 & -0.17 &  0.14 \\
SN2001ep & 10.30 &  0.72 &  0.48 &  0.80 &  0.93 \\
PTF10fps & 10.60 &  0.33 &  0.07 &  0.35 &  0.45 \\
SN2001ba & 11.20 &  0.35 &  0.06 &  0.16 &  0.21 \\
SN2001ep & 16.10 &  0.82 &  0.26 &  1.11 &  0.74 \\

\hline
\multicolumn{6}{c}{Undetermined$^{e}$}  \\
\hline
SN2009le &  0.30 & -0.35 & -0.05 & -0.81 &  0.16 \\
PTF10icb &  0.80 & -0.38 & -0.11 & -0.85 &  0.08 \\
PTF10bjs &  1.90 & -0.32 & -0.11 & -0.81 &  0.03 \\
PTF10pdf &  2.20 & -0.19 &  0.01 & -1.03 & -0.22 \\
PTF10acdh &  9.10 &  0.02 & -0.15 & -0.21 &  0.02 \\
\hline
\end{tabular}

\begin{tabular}{l}
$^{a}$$u-v$ spectrophotometry with $v$ filter response truncated at 5500\AA \\
and spectrophotometry shifted by 0.7 mag.  \\
$^{b}$Residual of $u-v$ spectrophotometry relative to linear best-fit. \\
$^{c}$At epochs of +5 days or later, NUV-blue and Irregulars cannot be distinguished. \\
$^{d}$ PTF10mwb=SN~2010gn \\
$^{e}$The grouping is considered undetermined for mixed $\Delta$ values. \\
\end{tabular}
\label{specphot_resids}
\end{table*}

\begin{table*}
\scriptsize
\vspace{-9mm}
\caption{Optical Parameters for the SN Ia Sample}
\begin{tabular}{l|c|cc|ccc}
\hline
\hline
SN & $\Delta m_{15}(B)$ & Carbon$^{a}$ & Ref$^{b}$ & HVG/LVG$^{c}$ &  HV/NV$^{c}$ & 
Ref.$^{b}$ \\
Name & [mag]            &  &    &          &      &       \\
\hline
\multicolumn{7}{c}{NUV-Blue}  \\
\hline
2006dd & 1.34 & A & 4 & --- & --- & --- \\
2008Q  & 1.40 & Y & 3 &  LVG & --- & 2 \\
2008hv & 0.95 & Y,A & 3,4 & LVG & --- & ---  \\
snf080514 & 1.20 & Y & 3 & LVG & --- & ---\\
2009dc$^{d}$ & 0.72 & Y,A & 2,3,8 & LVG & --- & 5 \\
2011by & 1.14 & A & 10 & LVG & NV &  13 \\
2011fe & 1.21 & A & 9 & LVG & NV &  11 \\

\hline
\multicolumn{7}{c}{NUV-Red} \\
\hline
2005cf & 1.07 & Y,A & 3,5 & LVG & NV & 1,2 \\
2005df & 1.20 & --- & & LVG & -- & 12 \\
2007af & 1.22 & N & 5   & LVG & NV &  1,6 \\
2007co & 1.09 & N & 5 & HVG & HV & 1,6 \\
2007cv & 1.31 & --- &  & --- & -- & --- \\  
2007gi & 1.37 & ?,N & 2,5 & HVG & HV & 1,2,5,6 \\
2007sr & 1.16 & --- & & HVG & -- & 6 \\ 
2008ec & 1.08 & N & 5 & LVG & NV & 5 \\
2009an & 1.20 & N & 14 &--- & --- & --- \\

\hline
\multicolumn{7}{c}{MUV-Blue} \\
\hline
2006dm & 1.54 & --- & --- & --- & NV & 6 \\   
2006ej & 1.39 & N & 5 & HVG & NV & 1,5 \\
2007cq & 1.04 & A & 5 & LVG & ---  & 5 \\
2010gn$^{e}$ & 1.19 & --- & --- &--- & --- & --- \\

\hline
\multicolumn{7}{c}{Irregular} \\
\hline
2009ig & 0.70 & P & 2 & HVG & HV & 2,6,7 \\
2009cz & 0.99 & N & 4 & --- & NV & 10 \\
10icb  & 0.80$^{f}$ &  F & 10 & --- & NV & 10 \\
\hline
\end{tabular}

\begin{tabular}{l}
$^{a}$ A=absorption line detected, F=flat-profile, N=not detected, Y=A/F not specified, \\
P=probable, ?=uncertain. \\
$^{b}$ References: (1) Wang et al. (2009b), (2) Parrent et al. (2011), (3) Thomas et al. (2011), \\
(4) Folatelli et al. (2012), (5) Silverman, Kong, \& Filippenko (2011),   
(6) X. Wang, private communication, \\
(7) Foley et al. (2011), (8) Silverman et al. (2011), (9) Nugent et al. (2011), \\
(10) J. Silverman, private communication, (11) Pereira et al. 2013, (12) X. Wang, private communication, \\
(13) Silverman, Ganeshalingam \& Filippenko 2013, (14) Determined by eye from CfA spectrum,  \\
(http://www.cfa.harvard.edu/supernova//spectra/sn2009an\_comp.gif) \\
$^{c}$ HVG/LVG refer to the velocity gradient of the SiII $\lambda$6355\AA absorption feature \\
(Benetti et al. (2005) and HV/NV refer to the velocity at peak of that line Wang et al. (2009b). \\
$^{d}$ SN~2009dc is a super-Chandrasekhar mass candidate, for which the HVG/LVG and HV/NV \\
classifications are not normally calculated. \\
$^{e}$ SN~2010gn=PTF10mwb. \\
$^{f}$ The peak width is derived from UVOT $b$ photometry. The PTF peak width, from \\
Maguire et al. (2012) derived from g \& r band data, is 1.09. \\  
\hline
\end{tabular}
\label{opt_params}
\end{table*}

\appendix
\renewcommand{\thetable}{A-\arabic{table}}
\setcounter{table}{0}
\section*{APPENDIX A: UVOT Photometry for 7 SNe~Ia}


Here we present UVOT photometry for seven SNe~Ia . 
The underlying galaxy light has been subtracted
(Brown et al. 2009) and the magnitudes
calibrated to the UVOT Vega system (Poole et al. 2008) 
using a time-dependent sensitivity and updated UV zeropoints from
Breeveld et al. (2011).

\begin{table}
\scriptsize
\caption{UVOT Photometry of Seven SNe Ia}
\begin{tabular}{lcccccc}
\hline
\multicolumn{7}{c}{SN 2009an} \\
\hline
\hline
JD$-$2,450,000 & $uvw2$ & $uvm2$ & $uvw1$ & $u$ & $b$ & $v$ \\
$[$days$]$ & [mag]$^{a}$ & [mag]$^{a}$ & [mag]$^{a}$ & [mag]$^{a}$ & [mag]$^{a}$
 & [mag]$^{a}$ \\
\hline
4893.67 & 17.91(06) & 19.80(23) & 16.30(05) & 14.56(04) & 14.73(04) & 14.75(04) \\
4894.17 & 17.87(09) & ---       & ---       &       --- &       --- &       --- \\
4896.84 & 17.78(06) & 19.14(10) & 16.23(04) & 14.55(04) & 14.64(04) & 14.67(04)  \\
4900.97 & 17.89(13) & 18.76(12) & 16.34(05) &       --- &       --- &       ---  \\
4902.24 & 18.05(07) & 19.03(16) & 16.44(06) &       --- &       --- &       ---  \\
4906.44 & 18.46(08) & 19.06(11) & 17.00(05) & 15.42(04) & 15.08(04) & 14.70(04)  \\
4911.93 & 18.92(09) & 19.43(12) & 17.64(07) & 16.20(05) & 15.70(04) & 15.00(04)  \\
4915.37 & 19.39(18) & 19.90(22) & 17.86(09) & 16.68(07) & 16.14(05) & 15.28(05)  \\
\\
\hline
\multicolumn{7}{c}{SN 2009cz} \\
\hline
\hline
4936.27 & 19.18(14) & ---       & 17.80(09) & 15.88(06) & 16.15(06) & 16.28(09) \\
4937.93 & 19.39(09) & ---       & 17.56(07) & 15.72(04) & 15.94(05) & 16.08(07) \\
4939.62 & 19.20(09) & 20.39(26) & 17.44(07) & 15.59(05) & 15.91(05) & 15.95(07) \\
4942.20 & 18.85(07) & 20.68(32) & 17.45(07) & 15.57(04) & 15.77(04) & 15.81(06) \\
4943.64 & 19.15(08) & ---       & 17.40(07) & 15.59(04) & 15.76(05) & 15.82(06) \\
4945.68 & 19.25(09) & 20.73(34) & 17.60(08) & 15.78(05) & 15.83(05) & 15.81(06) \\
4947.76 & 19.37(09) & 20.64(31) & 17.76(09) & 15.97(05) & 15.90(05) & 15.80(06) \\
4949.64 & 19.32(09) & ---       & 17.91(09) & 16.15(05) & 15.96(05) & 15.81(06) \\
4951.76 & 19.72(10) & ---       & 18.16(10) & 16.41(05) & 16.12(05) & 16.07(07) \\
4955.79 & 20.16(10) & ---       & 18.51(10) & 16.91(07) & 16.45(06) & 16.20(07) \\
4959.14 & 20.22(11) & ---       & 18.76(12) & 17.36(09) & 16.83(07) & 16.40(08) \\
4963.19 & 20.46(12) & ---       & 19.38(18) & 17.89(11) & 17.23(08) & 16.66(08) \\
4975.03 & ---       & ---       & 20.06(30) & 19.11(22) & 18.29(11) & 17.23(10) \\

\\
\hline
\multicolumn{7}{c}{PTF09dnl} \\
\hline
\hline
5071.57 & 19.29(17) & --- & 17.38(06) & 15.74(06) & 16.03(05) & 16.10(05) \\
5074.45 & 19.20(15) & --- & 17.34(06) & 15.75(06) & 16.00(04) & 16.00(05) \\
5076.33 & 19.38(17) & --- & 17.43(06) & 15.89(07) & 16.00(04) & 15.92(05) \\
5078.41 & 19.49(19) & --- & 17.66(07) & 16.13(07) & 16.06(04) & 15.96(05) \\
5089.35 & 20.37(32) & --- & 18.89(13) & 17.47(08) & 17.00(07) & 16.49(07) \\
\\
\hline
\end{tabular}
\begin{tabular}{l}
$^{a}$Uncertainties are in units of 0.01 mag. \\
\end{tabular}
\end{table}

\pagebreak
\begin{table}
\scriptsize
\caption{UVOT Photometry of Seven SNe Ia (continued)}
\begin{tabular}{lcccccc}
\hline
\multicolumn{7}{c}{PTF 10icb} \\
\hline
\hline
JD$-$2,450,000 & $uvw2$ & $uvm2$ & $uvw1$ & $u$ & $b$ & $v$ \\
$[$days$]$ & [mag]$^{a}$ & [mag]$^{a}$ & [mag]$^{a}$ & [mag]$^{a}$ & [mag]$^{a}$
 & [mag]$^{a}$ \\
\hline
5355.47 &   ---     &   ---     & 15.90(05) &    ---    &   ---     &    ---   \\
5356.47 &    ---     &   ---    &    ---   &  14.14(03) &   ---     &   ---    \\
5358.48 & 17.30(11) & 18.59(30) & 15.66(05) & 14.09(04) & 14.55(04) & 14.54(04) \\
5361.16 & 17.51(14) & 18.44(27) & 15.80(06) & 14.13(04) & 14.56(04) & 14.50(05) \\
5362.32 & 17.65(20) & 18.20(29) & 15.93(06) & 14.25(04) & 14.55(04) & 14.49(06) \\
5364.50 & 17.90(21) & 18.78(34) & 16.05(06) & 14.42(04) & 14.60(04) & 14.52(04) \\
5366.51 & 17.88(21) & 18.21(27) & 16.27(07) & 14.68(04) & 14.68(04) & 14.51(04) \\
5368.55 & 18.22(30) & 18.65(30) & 16.61(09) & 14.87(05) & 14.82(04) & 14.60(05) \\
5370.13 & 18.15(27) & 18.83(34) & 16.72(09) & 15.09(05) & 14.94(04) & 14.68(05) \\
5372.93 & 18.21(28) &  ---      & 17.15(13) & 15.40(06) & 15.15(05) & 14.80(05) \\
5374.16 &    ---    &   ---     & 17.26(13) & 15.57(06) & 15.26(05) & 14.90(05) \\
5376.33 &    ---    &   ---     & 17.46(15) & 15.90(07) & 15.50(05) & 15.05(06) \\
5378.08 & 18.63(32) &   ---     & 17.66(16) & 16.07(08) & 15.69(06) & 15.14(06) \\
5380.16 &    ---    &   ---     & 18.03(21) & 16.44(10) & 15.86(06) & 15.28(06) \\
\\
\hline
\multicolumn{7}{c}{SN 2010cr$^{b}$} \\
\hline
\hline
5310.65 & 19.70(20) & ---       & 18.50(14) & 16.52(06) &  16.78(06) & 16.89(11) \\
5313.08 & 19.65(18) & 20.02(23) & 18.46(13) & 16.86(08) &  16.86(07) & 16.78(09) \\
5315.68 & 20.27(19) & 20.30(19) & 18.97(13) & 17.29(08) &  16.94(06) & 16.97(08) \\
5317.62 & 20.34(20) & 20.83(26) & 18.97(13) & 17.70(09) &  17.23(07) & 16.96(08) \\
5319.69 & 20.58(25) & 20.65(24) & 19.39(16) & 18.04(11) &  17.58(07) & 17.09(08) \\
5321.16 & 20.70(28) & 20.85(30) & 19.97(25) & 18.20(11) &  17.78(07) & 17.21(09) \\
5323.50 & 20.76(27) & 21.15(34) & 20.34(31) & 18.57(14) &  18.25(09) & 17.34(09) \\
5325.12 & ---       & ---       & 20.45(34) & 19.09(19) &  18.63(11) & 17.50(10) \\
5327.35 & ---       & ---       & ---       & 19.84(35) &  18.96(14) & 17.73(12) \\
5331.90 & ---       & ---       & ---       & 19.76(33) &  19.38(19) & 18.31(18) \\
5333.78 & 21.02(31) & ---       & 20.32(29) & ---       &  19.41(18) & 18.39(18) \\
5344.46 & ---       & ---       & ---       & 20.74(35) &  20.25(25) & 18.79(27)  \\
5354.77 & ---       & ---       & ---       & ---       & ---        & 19.23(29) \\
\hline
\end{tabular}
\begin{tabular}{l}
$^{a}$Uncertainties are in units of 0.01 mag. \\
$^{b}$PTF10fps=SN 2010cr. \\
\end{tabular}
\end{table}

\pagebreak

\begin{table}
\scriptsize
\caption{UVOT Photometry of Seven SNe Ia (continued)}
\begin{tabular}{lcccccc}
\hline
\multicolumn{7}{c}{SN~2010gn$^{b}$} \\
\hline
\hline
JD$-$2,450,000 & $uvw2$ & $uvm2$ & $uvw1$ & $u$ & $b$ & $v$ \\
$[$days$]$ & [mag]$^{a}$ & [mag]$^{a}$ & [mag]$^{a}$ & [mag]$^{a}$ & [mag]$^{a}$
 & [mag]$^{a}$ \\
\hline
5385.21 & 20.14(33) & [20.28(04)] & 18.37(12) & 16.69(06) & 17.14(07) & 17.40(12) \\
5387.57 & 19.92(28) & [19.98(03)] & 18.28(11) & 16.55(05) & 16.94(06) & 17.20(11) \\
5391.47 & 19.92(28) & [19.94(04)] & 18.57(13) & 16.72(06) & 16.88(06) & 16.98(10) \\
5393.65 & 19.98(29) & [19.96(04)] & 18.62(13) & 16.93(06) & 16.94(06) & 17.06(10) \\
5396.84 & ----      & [20.04(02)] & 18.95(16) & 17.29(08) & 17.05(06) & 17.15(10) \\
5397.78 & ----      & [19.93(04)] & 19.07(18) & 17.37(08) & 17.11(06) & 17.22(11) \\
5399.58 & ----      & [20.04(01)] & 19.30(21) & 17.79(10) & 17.35(07) & 17.33(12) \\
5401.60 & ----      & [19.62(01)] & 19.72(29) & 18.14(13) & 17.61(08) & 17.13(14) \\
5403.63 & ----      & [19.93(02)] & 19.76(29) & 18.47(14) & 17.82(08) & 17.46(14) \\
5405.33 & ----      & [20.02(02)] & ----      & 18.49(15) & 18.02(09) & 17.70(15) \\
5407.34 & ----      & [20.06(01)] & ----      & 18.88(19) & 18.34(11) & 17.68(14) \\
\\
\hline
\multicolumn{7}{c}{SN 2011iv} \\
\hline
\hline
5900.25 & 15.29(04) & ---       & 13.92(04) & ---       & ---       & ---       \\
5901.41 & 15.18(04) & ---       & 13.82(04) & ---       & ---       & ---       \\
5901.84 & 15.15(05) & ---       & 13.82(04) & ---       & ---       & ---       \\
5902.48 & 15.14(04) & 15.53(05) & 13.80(04) & ---       & ---       & ---       \\
5903.34 & 15.16(04) & 15.50(04) & 13.81(04) & 12.29(03) & 12.79(03) & 12.73(04) \\
5904.79 & ---       & ---       & 13.89(04) & ---       & ---       & ---       \\
5905.72 & 15.27(05) & ---       & ---       & 12.43(03) & ---       &  ---      \\
5906.57 & 15.32(04) & 15.64(04) & 14.05(04) & ---       & ---       & ---       \\
5907.46 & 15.42(04) & ---       & ---       & 12.62(03) & 12.78(03) & 12.59(04) \\
5909.30 & 15.65(04) & 15.98(05) & 14.39(04) & 12.89(03) & 12.87(03) & 12.59(04) \\
5911.48 & 16.01(05) & 16.41(06) & 14.78(04) & 13.16(04) & 13.03(03) & 12.64(04) \\
5913.32 & 16.37(06) & 16.79(08) & 15.09(05) & 13.60(04) & 13.31(04) & 12.72(04) \\
5915.71 & 16.77(07) & 17.28(10) & 15.49(06) & 14.00(04) & 13.57(04) & 12.93(04) \\
5917.88 & 17.15(09) & 17.63(13) & 15.97(08) & 14.41(05) & 13.89(04) & 13.07(04) \\
5919.83 & 17.36(11) & 17.91(14) & 16.29(09) & 14.69(06) & 14.08(04) & 13.18(04) \\
5922.67 & 17.68(13) & 18.06(15) & 16.55(11) & 15.12(07) & 14.39(05) & 13.31(05) \\
5925.51 & 17.90(16) & 18.37(17) & 16.81(14) & 15.37(08) & 14.55(05) & 13.51(05) \\
5928.65 & 18.06(17) & 18.37(17) & 17.09(14) & 15.55(09) & 14.77(06) & 13.72(06) \\
5931.62 & 18.36(19) & 18.48(18) & 17.11(14) & 15.59(09) & 14.88(06) & 13.89(06) \\
5934.16 & 18.45(19) & 18.45(18) & 17.14(14) & ---       & ---       & ---       \\
\hline
\end{tabular}
\begin{tabular}{l}
$^{a}$Uncertainties are in units of 0.01 mag. \\
$^{b}$PTF10mwb=SN 2010gn. \\
\end{tabular}
\end{table}

\appendix
\renewcommand{\thetable}{A-\arabic{table}}
\setcounter{table}{0}
\section*{APPENDIX B: K-Corrections for NUV-red, irregular \& MUV-blue SNe Ia}

$K$-corrections are estimated from spectra to account for the effects of 
redshift on the flux through a filter, as measured here on Earth. 
Although the small aperture of the 
UVOT instrument dictates low-redshift SNe Ia as targets, the dramatic drop in 
flux blueward of 4000\AA\ typical of SN Ia spectra suggests that $K$-corrections 
could even be important in this sample. The obvious challenge for applying $K$-corrections 
is to accumulate time sequences of UV spectra of all major and minor groups to 
determine accurate $K$-corrections for all SNe Ia. Recent HST and UVOT spectral campaigns 
are improving the situation (e.g. Foley et al. 2011, Foley et al. 2012), but to date, 
the spectral sampling of SNe Ia observed in the UV is: 1) not adequate for the generation of 
$K$-correction sequences in the $uvw1$, $uvw2$, $uvm2$-bands, 2) better in the $u$-band and 
3) unimportant in the $b$ and $v$-bands.
   
Our first effort to generate $K$-corrections was presented in B10, where  
$K$-corrections were obtained by ``mangling" three spectra
(e.g. a +5 day spectrum of SN~1992A (Kirshner et al. 1993), a maximum-light
template from Hsiao et al. (2009) and a maximum-light
template from Nugent et al. (2002), adjusting the wavelength of each spectrum for the measured
redshift of each SN, and then convolving that spectrum with the UVOT filter
transmissions. The Hsiao template included rest-frame UV spectra from high-$z$ 
SNe Ia, while the Nugent template sequence added $IUE$ spectra of many SNe Ia to 
the SN~1992A spectra. Mangling starts with a standard spectrum and distorts it to create 
spectrophotometry that matches the observed photometry. As a cross-check, B10 
compared the $K$-corrections from the three starting spectra, finding fairly small 
differences between the three estimates of the $K$-corrections. The sample in 
B10 included one SN Ia that we would now term NUV-blue (2008Q) and two that would now 
be termed MUV-blue (2007cq \& 2006ej). 
The mangling of the spectra will add emission to account for the NUV-optical
excess for the NUV-blue sample, but it is not clear whether the resulting spectral
shape is in any way related to the unobserved, true spectral shape. 

In this appendix, we will investigate only $K$-corrections for the $u$-band as calculated 
from the available spectra and from the spectral mangling employed in B10. 
B10 determined that the K$_{b}$ and K$_{v}$ corrections were typically 
0.01$^{m}$ -- 0.02$^{m}$, with a maximum of 0.04$^{m}$. For the current application 
of color curves and absolute magnitudes, we
consider those values negligible, and since the $b$ and $v$ colors of all major and 
minor groups are similar, we do not further address K$_{b}$ and K$_{v}$.
The $K$-corrections reported in B10 were not negligible for the filters bluer than 
the $b$ filter. The mean corrections reported in B10 are 0.09$^{m}$, 0.15$^{m}$ and 0.06$^{m}$ 
for K$_{uvm2}$, K$_{uvw1rc}$ and K$_{u}$, respectively. In order to apply $K$-corrections for the 
$uvw1$, $uvw2$, $uvm2$-bands, the spectra must extend blueward to $\sim$1600 \AA\ and 
redward to at least 3500 \AA\, as dictated by the spectral transmission curves of the UVOT filters. 
Few observed UV spectra meet those criteria combined with a UVOT color-curve grouping determinations 
(as reported in this work), meaning that checking the $K$-corrections from spectral mangling versus 
corrections from actual spectra is not possible for the bluest filters. In this current work, we 
ignore $K$-corrections in the $uvw1$, $uvw2$, $uvm2$-bands, rather than employ the corrections from 
spectral mangling. This decision was based upon the importance placed on avoiding biasing the 
presentation of color differences in this initial presentation of these differences. 
A future effort will concentrate on trying to derive 
$K$-corrections for all major and minor groups from high-$z$ SNe Ia, where the rest-frame UV is 
observed in the optical. That effort hinges on whether there can be unambiguous determination of 
the SN Ia grouping from the available optical observations. Obtaining spectra from low-$z$ SNe Ia 
that reaches 1800 \AA\ would also be valuable.   

\begin{figure}
\epsscale{1.1} \plotone{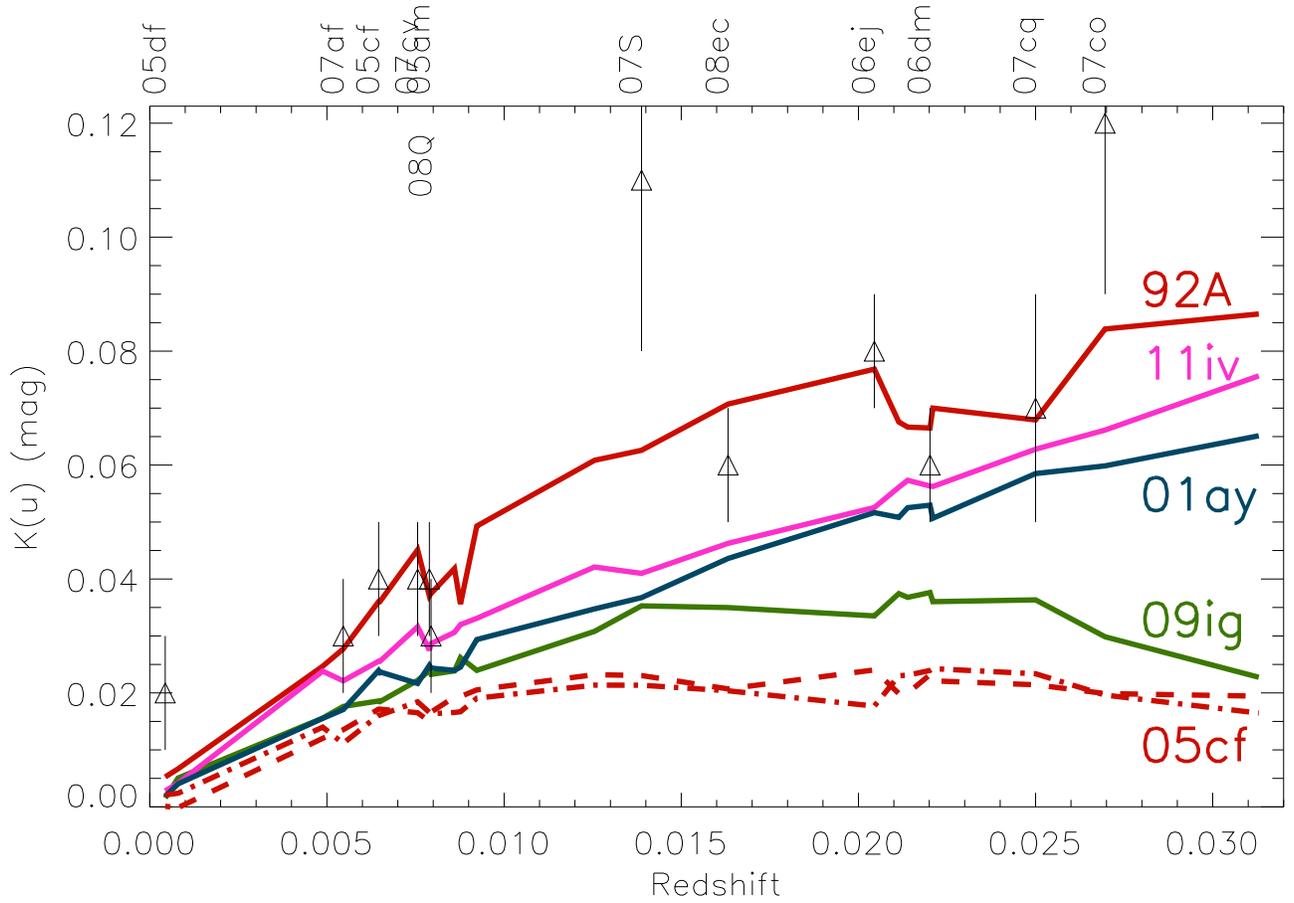}
\caption{Near-peak K-corrections for the $u$-band based upon three spectra. 
NUV-red corrections are shown in red and are  
based on an HST/optical spectrum of SN~1992A (Kirshner et al. 1993) (solid-line) and two 
spectra of SN~2005cf (Wang et al. 2009a) (dashed lines). 
NUV-red-irregular corrections are shown in green and
are based on a {\it Swift} UVOT spectrum of SN~2009ig.
Narrow-peaked corrections are shown in pink and are based on an HST/optical spectrum 
of SN~2011iv (Foley et al. 2012b). 
The $K$-corrections presented in B10 
are shown as open diamonds, and in most cases agree with the SN~1992A corrections derived in this work.}
\label{kcorru}
\end{figure}

Far more spectra span the 3000 \AA\ -- 4000 \AA\ wavelength range important for determining 
K$_{u}$. Figure \ref{kcorru} shows K$_{u}$ for a range of redshifts as calculated from 
near optical peak 
spectra of 5 SNe~Ia. The NUV-red spectra are a SN~1992A composite spectrum from Kirshner et al. 1993, 
and two SN~2005cf composite spectra at -0.8$^{d}$ and +2.2$^{d}$, from Wang et al. 2011. 
The narrow-peaked spectrum is a composite spectrum of SN~2011iv at +0.6$^{d}$ from Foley at al. 2012a. 
The NUV-red-irregular spectrum is the average of two {\it Swift} spectra at -4.2$^{d}$ and -2.1$^{d}$ 
from Foley et al. 2012b. The NUV-blue spectrum is a spectrum of SN~2001ay 
from Foley, Filippenko \& Jha 2008. The derived K$_{u}$ values agree with values from B10 for the 
majority of SNe Ia studied. Most importantly, the corrections range from 0.02$^{m}$ -- 0.08$^{m}$, 
suggesting that ignoring the color groupings will induce an error of 0.06$^{m}$ or less for 
SNe~Ia beyond z=0.03. There is no clear distinction between the groups, as the NUV-red events 
have both the largest and smallest corrections in this sample. 
The lack of clear distinctions between the groups relative to the variations within each group 
leads us to not employ K$_{u}$ corrections in this work.

\begin{thebibliography}{}
\singlespace
\bibitem[Altavilla et al. (2007)]{Altavilla_etal_2007}Altavilla, G., et al. 2007, A\&A, 475, 585
\bibitem[Benetti et al. (2005)]{Benetti_etal_2005}Benetti, S., et al. 2005,
   ApJ, 623, 1011 
\bibitem[Blondin et al. (2012)]{Blondin_etal_2012}Blondin, S., et al. 2012, AJ, 143, 126
\bibitem[Breeveld et al. (2011)]{Breeveld_etal_2011}Breeveld, A. A., et al. 2011,
in American Institute of Physics Conference Series, Vol. 1358, ed. J. E. McEnery, 
J. L. Racusin \& N. Gehrels, 373-376 
\bibitem[Brown et al. (2009)]{Brown_etal_2009}Brown, P. J., et al. 2009, 
AJ, 137, 4517
\bibitem[Brown et al. (2010)]{Brown_etal_2010}Brown, P. J., et al. 2010, 
AJ, 721, 1608 (B10)
\bibitem[Brown et al. (2012a)]{Brown_etal_2012a}Brown, P. J., et al. 2012a,
ApJ, 749, 18  
\bibitem[Brown et al. (2012b)]{Brown_etal_2012b}Brown, P. J., et al. 2012b, 
ApJ, 753, 22
\bibitem[Bufano et al. (2009)]{Bufano_etal_2009}Bufano, F., et al. 2009, 
ApJ, 700, 1456
\bibitem[Cappellaro et al. (1995)]{Cappellaro_etal_1995}Cappellaro, E., Turatto, M. 
\& Femley J. 1995, IUE-ULDA Access Guide 6: Supernovae (ESA-SP 1189; 
Noordwijk: ESA) 
\bibitem[Cooke et al. (2011)]{cooke11}Cooke, J., et al. 2011, ApJ, 727, 35
\bibitem[Ellis et al. (2008)]{ellis08}Ellis, R.S., et al. 2008, ApJ, 674, 51
\bibitem[Foley, Filippenko \& Jha (2008)]{Foley_etal_2008}Foley, R. J., Filippenko, A. V., 
  \& Jha, S.W. 2008, ApJ, 686, 467   
\bibitem[Foley \& Kasen (2011)]{foley_Kasen_2011}Foley, R. J. \& Kasen, D. 2011, ApJ, 729, 55
\bibitem[Foley et al. (2011)]{foley11}Foley, R. J., et al. 2011, ApJ, 742, 89
\bibitem[Foley et al. (2012a)]{foley12a}Foley, R. J., et al. 2012a, ApJ, 744, 38
\bibitem[Foley et al. (2012b)]{foley12b}Foley, R. J., et al. 2012b, ApJ, 753, L5
\bibitem[Foley et al. (2012c)]{foley12c}Foley, R. J., et al. 2012c, AJ, 143, 113
\bibitem[Foley \& Kirshner (2013)]{foley_kirshner2013}Foley, R. J. \& Kirshner, R.P. 
2013, ApJ, 753, 5  
\bibitem[Folatelli et al. (2010)]{folatelli10}Folatelli, G., et al. 2010,
   AJ, 139, 120
\bibitem[Folatelli et al. (2012)]{folatelli12}Folatelli, G., et al. 2012,
   ApJ, 745, 74
\bibitem[Ganeshalingam et al. (2010)]{Ganeshalingam_etal_2010}Ganeshalingam, M., et al. 2010, 
ApJS, 190, 418
\bibitem[Gehrels et al. (2004)]{Gehrels_etal_2004} Gehrels, N., et al. 2004, 
\apj, 611, 1005
\bibitem[Hachinger et al. (2013)]{hach13}Hachinger, S., et al. 2013, MNRAS, 429, 2228
\bibitem[Hamuy et al. (1996a)]{Hamuy_etal_1996abs}Hamuy, M., Phillips, M. M., 
Suntzeff, N. B., Schommer, R. A., Maza, J., \& Aviles, R. 1996, AJ, 112, 2391
\bibitem[Hamuy et al. (1996b)]{Hamuy_etal_1996shapes}Hamuy, M., Phillips, M. 
M., Suntzeff, N. B., Schommer, R. A., Maza, J., Smith, R. C., Lira, P., \&
Aviles, R. 1996, AJ, 112, 2438
\bibitem[Hicken et al. (2009)]{Hicken_etal_2009}Hicken, M., Wood-Vasey, W.M., Blondin, S., 
Challis, P., Jha, S., Kelly, P., Rest, A., Kirshner, R.P 2009, ApJ, 700, 1097
\bibitem[H\"{o}flich et al. (2010)]{Hoflich_etal_2010}H\"{o}flich, P., et al. 2010, ApJ, 710, 444
\bibitem[Hsiao et al. (2009)]{hsiao_etal_2009}Hsiao, E.Y., et al. 2009, \apj, 663, 1187
\bibitem[Iwamoto et al. (1999)]{iwam99} Iwamoto, K., et~al. 1999, ApJS, 125, 439
\bibitem[Jeffery et al. (1992)]{jeff92}Jeffery, D. J., Leibundgut, B., 
Kirshner, R. P., Benetti, S., Branch, D., \& Sonneborn, G. 1992, ApJ, 
397, 304
\bibitem[Jha, Riess \& Kirshner (2007)]{JhaRiessKirshner}Jha, S., Riess, A. \& 
Kirshner, R.P. 2007, ApJ, 659, 122 
\bibitem[Kirshner et al. (1993)]{Kirshner_etal_1993} Kirshner, R. P., et 
  al. 1993, ApJ, 415, 589
\bibitem[Krisciunas et al. (2011)]{Kris_etal_2011} Krisciunas, K. et al. 2011, 
AJ, 142, 74  
\bibitem[Lentz et al. (2000)]{Lentz_etal_2000}Lentz, E. J., Baron, E.,
Branch, D., Haushildt, P. H., \& Nugent, P. E. 2000, ApJ, 530, 966
\bibitem[Lira (1995)]{Lira_1995}Lira, P. 1995, Masters thesis, Univ. Chile
\bibitem[Maguire et al. (2012)]{Maguire_etal_2012}Maguire, K. et al. 2012, MNRAS, 426, 2359
\bibitem[Mazzali (2000)]{Mazzali_2000}Mazzali, P.A. 2000, A\&A, 363, 705
\bibitem[Mazzali et al. (2005)]{Mazzali_etal_2005}Mazzali, P.A., et al. 2005, 
ApJ, 623, L37
\bibitem[Milne et al. (2010)]{milne10}Milne, P.A. et al. 2010, AJ, 721, 1627 (M10)
\bibitem[Nugent et al. (2002)]{nugent_etal_2002}Nugent, P., Kim, A., \& Perlmutter, S. 2002, 
PASP, 114, 803 
\bibitem[Nugent et al. (2011)]{nugent_etal_2011}Nugent, P., et~al. 2011, Nature, 480, 344
\bibitem[Panagia (2003)]{Panagia_2003}Panagia, N. 2003, in Lecture Notes in Physics 598, 
Supernovae and Gamma-Ray Bursters, ed. K. Weiler (Berlin: Springer), 113
\bibitem[Parrent et al. (2011)]{parrent11}Parrent, J. et~al. 2011, 
ApJ, 732, 30
\bibitem[Patat et al. (2012)]{Patet_etal_2012}Patat, F., H\"{o}flich, P., Baade, D., Maund, J. R., 
Wang, L., Wheeler, J. C. 2012, A\&A, 545, A7
\bibitem[Perlmutter et al. (1997)]{Perlmutter_etal_1997}Perlmutter, S., et al. 1997, APJ, 483, 565
\bibitem[Perlmutter et al. (1999)]{Perlmutter_etal_1999}Perlmutter, S., et 
  al. 1999, ApJ, 517, 565
\bibitem[Phillips (1993)]{Phillips_1993}Phillips, M. M. 1993, ApJ, 413, L105
\bibitem[Phillips et al. (1999)]{Phillips_etal_1999}Phillips, M. M., Lira, P.,
     Suntzeff, N. B., Schommer, R. A., Hamuy, M., \& Maza, J. 1999,
     AJ, 118, 1766 
\bibitem[Poole et al. (2008)]{Poole_etal_2008}Poole, T., et al. 2008, 
MNRAS, 383, 627
\bibitem[Richmond \& Smith (2012)]{Richmond_Smith_2012}Richmond, M. W. \& 
Smith, H. A. 2012, JAVSO, 40, 872
\bibitem[Riess, Press \& Kirshner (1996)]{Riess_96MLCS}Riess, A. G., 
Press, W. H., \& Kirshner, R. P. 1996, ApJ, 473, 88
\bibitem[Riess et al. (1998)]{Riess_etal_1998}Riess, A. G., et al. 1998, 
AJ, 116, 1009
\bibitem[Roming et al. (2000)]{Roming_etal_2000}Roming, P.W.A, et al. 2000, 
SPIE, 4140, 76
\bibitem[UVOT: Roming et al. (2005)]{Roming_etal_2005}Roming, P. W. A., 
et al. 2005, Space Science Reviews, 120, 95
\bibitem[Sauer et al. (2008)]{Sauer_etal_2008}Sauer, D. N., Mazzali, P. A., 
Blondin, S., Filippenko, A. V., Benetti, S., Stehle, M., Challis, P., 
Kirshner, R. P., \& Li, W. 2008, MNRAS, 391, 1605
\bibitem[Schlafly \& Finkbeiner (2011)]{schl11}Schlafly \& Finkbeiner. {\it ApJ}, {\bf 737}, 103 (2011)
\bibitem[Silverman et al. (2011)]{silv11}Silverman,  J. M., Ganeshalingam, M.,
 Li, W., Filippenko, A.V., Miller, A.A., Poznanski, D. 2011, MNRAS, 410, 585
\bibitem[Silverman, Kong \& Filippenko (2012)]{Silverman_Kong_Filippenko_2012}Silverman, J.,
Kong, J.J. \& Filippenko, A.V. 2011, MNRAS, 425, 1819 
\bibitem[Silverman \& Filippenko (2012)]{Silverman_Filippenko_2012}Silverman, J. \& Filippenko, A.V. 2012,
MNRAS, 425, 1917  
\bibitem[Silverman, Ganeshalingam \& Filippenko (2013)]{Silverman_Ganeshalingam_Filippenko_2013}Silverman, J.,
Ganeshalingam, M., \& Filippenko, A.V. 2013, MNRAS, 430, 1030 
\bibitem[Stritzinger et al. (2010)]{stritz10}Stritzinger, M., et al. 2010, 
AJ, 140, 2036
\bibitem[Taubenberger et al. (2010)]{Taubenberger_etal_2010}Taubenberger, S., et al. 2010, 
MNRAS, 412, 2735
\bibitem[Thomas et al. (2011)]{Thomas_etal_2011}Thomas, R.C. et al. 2011, ApJ, 743, 27
\bibitem[Walker et al. (2012)]{Walker_etal_2012}Walker, E.S. et al. 2012, MNRAS, 427, 103
\bibitem[Wang et al. (2009a)]{Wang_etal_2009a}Wang, X., et al. 2009a, 
ApJ, 697, 380
\bibitem[Wang et al. (2009b)]{Wang_etal_2009b}Wang, X., et al. 2009b,
  ApJ, 699, 139   
\bibitem[Wang et al. (2012)]{Wang_etal_2012}Wang, X., et al. 2012, ApJ, 749, 126
\bibitem[Wang et al. (2013)]{Wang_etal_2013}Wang, X., Wang, L., Filippenko, A. V., 
Zhang, T., Zhao, X. 2013, Science, 340, 170
\end{thebibliography}
\end{document}